\documentclass[a4paper,12pt]{article}

\usepackage{xr}
\usepackage[utf8]{inputenc}
\usepackage{latexsym}
\usepackage{xcolor}
\usepackage{graphicx}
\usepackage{amsmath}
\usepackage{setspace}
\usepackage{lineno}
\usepackage{titling}
\usepackage{siunitx}
\usepackage{natbib}
\usepackage[hidelinks]{hyperref}
\usepackage{cleveref}
\usepackage{float}
\usepackage{authblk}
\usepackage{sectsty}

\usepackage{caption}
\subsectionfont{\itshape}

\title{How does thermal pressurization of pore fluids affect 3D strike-slip earthquake dynamics and ground motions?}

\author[1]{Jagdish Chandra Vyas}
\author[3,2]{Alice-Agnes Gabriel}
\author[2]{Thomas Ulrich}
\author[1]{Paul Martin Mai}
\author[4]{Jean-Paul Ampuero}
\affil[1]{\normalsize{King Abdullah University of Science and Technology, Saudi Arabia}}
\affil[2]{\normalsize{Department of Earth and Environmental Sciences} \\ \normalsize{Ludwig-Maximilians-Universit\"at M\"unchen, Munich, Germany}}
\affil[3]{\normalsize{Institute of Geophysics and Planetary Physics, Scripps Institution of Oceanography, University of California San Diego, La Jolla, USA}}
\affil[4]{Geoazur Laboratory, Université Côte d'Azur, France}
\date{October, 2022}

\begin{document}

\maketitle


\newpage
\begin{abstract}

Frictional heat during earthquake rupture raises the pressure of fault zone fluids and affects the rupture process and its seismic radiation. Here, we investigate the role of two key parameters governing thermal-pressurization of pore fluids -- hydraulic diffusivity and shear-zone half-width -- on earthquake rupture dynamics, kinematic source properties and ground-motions. We conduct 3D strike-slip dynamic rupture simulations assuming a rate-and-state dependent friction law with strong velocity-weakening coupled to thermal-pressurization of pore fluids. Dynamic rupture evolution and ground-shaking are densely evaluated across the fault and Earth surface to analyze variations of rupture parameters (slip, peak slip-rate PSR, rupture speed Vr, rise time Tr), correlations among rupture parameters, and variability of peak ground velocity (PGV).

Our simulations reveal how variations in thermal-pressurization affect source properties. We find that mean slip and Tr decrease with increasing hydraulic diffusivity, whereas mean Vr and PSR remain almost constant. Mean slip, PSR and Vr decrease with increasing shear-zone half-width, whereas mean Tr increases. Shear-zone half-width distinctly affects the correlation between rupture parameters, especially for parameter pairs slip-Vr, PSR-Vr and Vr-Tr. Hydraulic diffusivity has negligible effects on these correlations. Variations in shear-zone half-width primarily impact Vr, which then may affect other rupture parameters.  We find negative correlation between slip and PSR, in contrast to simpler dynamic rupture models, whereas trends for other parameter pairs are in agreement. Mean PGVs decrease faster with increasing shear-zone half-width than with hydraulic diffusivity, whereas ground-motion variability is similarly affected by both parameters. Our results show that shear-zone half-width affects rupture dynamics, kinematic rupture properties and ground-shaking  more strongly than hydraulic diffusivity. We interpret the importance of shear-zone half-width based on the characteristic time of diffusion.


\end{abstract}

\newpage
\section*{Introduction} 
\paragraph{} 
Most rocks in the upper crust are porous and are likely to contain small amounts of fluids, and fault zones often are privileged paths for fluid transport in the crust. Pore fluids can affect the stress and strain conditions of rocks before, during and after an earthquake \citep{sibson1973interactions, sibson1977kinetic, sibson1980power, lamb2006shear, madden_state_2022}. Frictional heat produced in the slip zone during earthquake rupture can cause thermal expansion, flow and pressurization of pore fluids impacting the stress conditions on the fault and in the surrounding rocks \citep{sibson1973interactions, sibson1980power, lachenbruch1980frictional, mase1984pore, mase1987effects, andrews2002fault}.
Thermal pressurization of pore fluids (TP) is considered a dominant dynamic weakening mechanism affecting earthquake rupture nucleation, propagation and arrest.
The TP mechanism operates as follows: thermal expansion and confinement of fluids within a rock matrix of lower thermal expansivity produces a stark increase of pore fluid pressure; this decreases effective normal stress, which in turn lowers fault frictional strength, thereby enhancing fault dynamic weakening during earthquake rupture \citep{andrews2002fault, suzuki2006nonlinear, bizzarri2006thermal_part1, bizzarri2006thermal_part2, noda2010three, garagash2012seismic, viesca_ubiquitous_2015, brantut2019stability, badt_thermal_2020}.
The resulting extreme reduction of fault strength may prevent co-seismic melting and resolve the so-called heat-flow paradox \citep{lachenbruch1980heat, andrews2002fault, di2011fault, acosta2018dynamic, badt_thermal_2020}.

\paragraph{}
TP effects may play a significant role in modulating the magnitude of induced or triggered earthquakes in georeservoirs, governed by pressure change and fluid flow \citep[e.g.][]{galis_induced_2017}. Fluid-saturated clay-rich fault materials present in subduction zone environments may promote earthquake rupture propagation and slip due to pressurisation of pore fluid and low permeability of surrounding materials, resulting in dynamic weakening behaviour and low breakdown work \citep{hirono2016near, aretusini2021fluid}. TP may also govern subduction dynamics, as suggested by geological signatures found in exhumed ancient subduction thrusts, including fluidization of comminuted material and increase in the volume of fluid inclusions by frictional heating \citep{ujiie_geological_2010}. TP effects may lead to slip-weakening distances ranging from 0.03 to 0.22 m for subduction thrusts, which indicates high radiated energy and fast stress release \citep{ujiie_geological_2010}. TP may also explain why stable creeping parts of the fault become unstable during coseismic slip, as observed during the 2011 $M_w \, 9.0$ Tohoku Oki earthquake  \citep{mitsui_scenario_2012, noda_stable_2013}. 

\paragraph{} 
Theoretical and numerical studies suggest that TP may also affect earthquake rupture properties such as stress drop and fracture energy.
\cite{andrews2002fault} shows that the stress drop associated with TP rises with increasing rupture propagation distance, and may eventually dominate over the frictional induced stress drop. 
Recently, \cite{perry2020nearly} demonstrated that enhanced dynamic weakening due to TP is consistent with the observationally inferred magnitude-invariance of stress drop. 
Simulations of earthquake sequences on rate-and-state faults with TP suggest that major faults operate at low overall shear stress levels, while the peak stress at the rupture front is consistent with static friction coefficients of $0.6 - 0.9$ \citep{noda2009earthquake}. 
\cite{viesca_ubiquitous_2015} hypothesize that TP may explain the observed distinct transition in scaling of fracture energy with earthquake size, which implies different fault weakening processes for large and small earthquakes. Moreover, the seismologically inferred increase in fracture energy with earthquake magnitude may also be explained with TP \citep{abercrombie2005can, rice2006heating, viesca_ubiquitous_2015, perry2020nearly}. 
In dynamic earthquake rupture simulations, TP facilitates rupture propagation and activation of fault branches, and increases rupture speed and slip rate \citep{urata2014effect, schmitt2015nucleation}. 

\paragraph{} 
\cite{lambert2021propagation} examined why large earthquakes occur at lower levels of stress than their expected static strength using numerical seismic-cycle modeling. They tested two competing hypotheses: (1) faults are quasi-statically strong but experience significant weakening during earthquake rupture, and (2) faults are persistently weak because of pore fluid overpressure. They found that the former produces sharper self-healing pulse-like ruptures, larger dynamic stress changes and larger radiated energy than that inferred for megathrust earthquakes, whereas the latter produces crack-like ruptures. Therefore, their findings suggest two possibilities: either large earthquakes do not propagate as self-healing pulses, or the radiated energy is substantially underestimated for megathrust earthquakes. Their results raise questions about earthquake rupture physics for large events and their weakening mechanisms.

\paragraph{}
\cite{acosta2018dynamic} performed laboratory experiments to investigate the role of flash heating and TP on dynamic weakening of faults. They observed that loss in fault strength is due to flash heating under dry and low pressure (1~MPa) conditions, whereas at high fluid pressure (25~MPa) flash heating is inhibited and TP becomes important.
\cite{badt_thermal_2020} tested the mechanical response of faults under elevated
confining and pore pressures in laboratory experiments and suggest that the magnitude and rate of weakening increases as the sample permeability decreases with limited frictional heat production. Their experiments support the hypothesis that TP is an active dynamic weakening mechanism during earthquake rupture at least in the early stages of fault slip.

\paragraph{} 
Seismic radiation and near-field ground-motion characteristics depend on earthquake rupture dynamics. 
Observations of strong ground motions show intricate variations depending on parameters like source-to-site distance, earthquake magnitude and faulting style, hypocenter position, azimuth with respect to fault plane and rupture direction, and more \citep{youngs1995magnitude, rodriguez2011analysis, imtiaz2015ground, vyas2016distance, gallovivc2017azimuthal, crempien2018within}.
The question is how TP may modulate the intensity and spectral properties of near-fault ground motions. For instance, a higher spectral velocity response at long periods ($\sim$ 10~s) observed at receivers near the northern portion of the main fault of the Mw 7.6 1999 Chi-chi, Taiwan earthquake, compared to recordings at receivers near the southern portion of fault, has been attributed to TP \citep{andrews2005thermal}. Hence, here we investigate the effect of TP on rupture dynamics and
ground-motion properties. 

\paragraph{} 
Several numerical implementations have been proposed to include TP in dynamic earthquake rupture modeling based on the theoretical foundations of the conservation of fluid mass and energy, Fourier’s law and Darcy’s law \citep{andrews2002fault, bizzarri2006thermal_part1, bizzarri2006thermal_part2, noda2009earthquake, noda2010three, schmitt2015nucleation}. 
However, no detailed investigation exists on how TP influences dynamic and kinematic properties and ground motion characteristics associated with earthquake rupture. In this study, we perform a parametric sensitivity analysis using 3D dynamic rupture modeling with a strong velocity-weakening rate-and-state friction law and TP to examine and quantify the influence of TP on kinematic rupture characteristics, such as fault slip, peak slip rate (PSR), rupture speed (Vr) and rise time (Tr). Correlation among these kinematic rupture parameters has been previously analyzed using numerical models that did not incorporate TP  \citep[e.g.,][]{oglesby2002stochastic, guatteri2003strong, schmedes2010correlation, schmedes2013kinematic, mai2018accounting}.
Here, we examine how TP alters such correlations. Additionally, we investigate the effects of TP on earthquake ground-motion properties. 

\section*{Computational method and model setup} 
\paragraph{}
We perform 3D dynamic rupture modeling of strike-slip earthquakes accounting for TP. The 3D benchmark parametrization by \citet{ulrich_TPV1053d} forms the basis for these simulations.
In this section, we provide details on the fault friction law, the TP model, the numerical method and the parameter space explored. 

\subsection*{Frictional properties} 
\begin{figure}[H]
    \centering
    \includegraphics[width=0.75\textwidth]{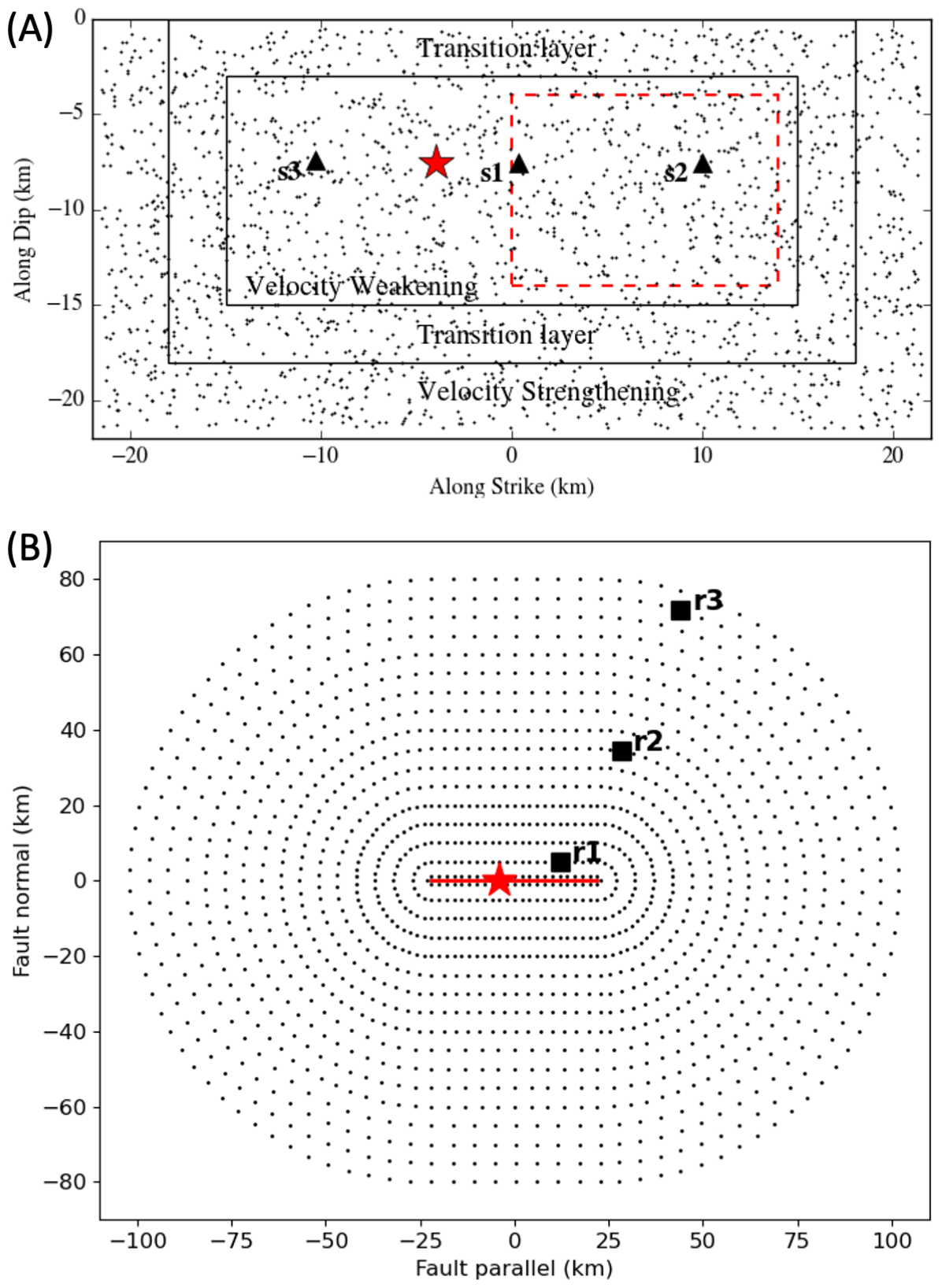}
    \caption{(A) On-fault receivers at which rupture parameters are evaluated. Black dots mark receivers used in the statistical analysis, the three triangles (s1, s2, s3) denote selected receivers at which we analyse the detailed temporal evolution of rupture parameters. Receivers inside the red dashed rectangle are used for analyzing rupture parameter correlations. (B) Receiver array at the Earth surface for ground-motion analysis. Black dots mark receivers used in the statistical analysis and the three squares (r1, r2, r3) locate receivers used for analysing ground velocity time-histories.}
    \label{fig:rec}
\end{figure}

\paragraph{}
We adopt a strong velocity-weakening rate-and-state friction law \citep{dunham2011earthquake} that replicates the large friction reduction observed in laboratory experiments at co-seismic slip rates \citep{di2011fault}.
The shear strength of the fault is the product of the friction coefficient $f$ and effective normal stress $\overline{\sigma}$: 
\begin{equation}
\label{tau}
    \tau=f(V,\Psi) \overline{\sigma}.
\end{equation}
The friction coefficient is a function of slip velocity ($V$) and a state variable ($\Psi$):
\begin{equation}
\label{fv}
    f(V,\Psi) = a \,\, \sinh^{-1}\left[\frac{V}{2V_0}  \, exp\left(\frac{\Psi}{a}\right)\right].
\end{equation}
The state variable evolves according to the following equation:
\begin{equation}
    \frac{d\Psi}{dt}=-\frac{V}{L}[\Psi-\Psi_{\textsc{ss}}(V)],
\end{equation}
where the steady-state value of the state variable is
\begin{equation}
    \Psi_{\textsc{ss}}(V)=a \, ln\left\{\frac{2V_0}{V}\sinh\left[\frac{f_{\textsc{ss}}(V)}{a}\right]\right\}.
\end{equation}
The steady state friction coefficient $f_\textsc{ss}$ depends on $V$ as
\begin{equation}
    f_{\textsc{ss}}(V)=f_w+\frac{f_{\textsc{lv}}(V)-f_w}{\left[1+(V/V_w)^8\right]^{1/8}},
\end{equation}
with a low-velocity steady-state friction coefficient given by
\begin{equation}
    f_{\textsc{lv}}(V)=f_0-(b-a) \, \ln(V/V_0).
\end{equation}
The reference friction coefficient ($f_0$), reference slip velocity ($V_0$), direct effect parameter ($a$), state evolution effect parameter ($b$), fully-weakened friction coefficient ($f_w$), weakening slip velocity ($V_w$) and state evolution distance ($L$) are chosen as constant over the central region of the fault, a rectangular velocity-weakening asperity (see Figure \ref{fig:rec} and Table \ref{tab:MP-FP-TP-IC}). The frictional behaviour transitions to velocity-strengthening in the surrounding region (Figure \ref{fig:rec}A), over a transition zone with a width ($h$) of 3 km, and allows a gradual arrest of the rupture in all simulations. Rupture is inhibited by increasing the direct effect parameter up to $a + \Delta a_0$ and the weakening slip velocity up to $V_w + \Delta V_{w0}$ via the following spatial distributions:

\begin{equation}\label{EQ1}
    \Delta a(x, z) = \Delta a_0 [1-B_1(x;H,h) \, B_2(z;H,h],
\end{equation}
\begin{equation}\label{EQ2}
    \Delta V_w(x, z) = \Delta V_\textsc{w0} [1-B_1(x;H,h) \, B_2(z;H,h)],
\end{equation}

where

\begin{equation}\label{EQ3}
B_1(x;H,h)=
    \begin{cases}
    1, \,\, |x|  \leq H 
    \\
    0.5 \left[1+\tanh\left(\frac{h}{|x|-H-h}+\frac{h}{|x|-H} \right) \right], \,\, H<|x|<H+h
    \\
    0, \,\, |x| \geq H+h,
    \end{cases}
\end{equation}

\begin{equation}\label{EQ4}
B_2(z;H,h)=
    \begin{cases}
    0.5 \left[1+\tanh\left(\frac{h}{h-z}-\frac{h}{z} \right) \right], \,\, z<h
    \\
    1, \,\, h \leq z \leq H
    \\
    0.5 \left[1+\tanh\left(\frac{h}{z-H-h}+\frac{h}{z-H} \right) \right], \,\, H<z<H+h
    \\
    0, \,\, z \geq H+h,
    \end{cases}
\end{equation}

$(x,y,z)$ defines a right-handed coordinate system in which $x$ is the strike direction and $z$ the vertical upwards direction, such that $x=0$ is in the center of the fault and $z=0$ at the ground surface. In this study, we assume $H$ = 15 \si{\km}, $h$ = 3 \si{\km}, $\Delta a_0$ = 0.01 and  $\Delta V_\textsc{w0}$ = 0.95.

\subsection*{Thermal pressurization}
\paragraph{}
To include TP effects, we solve the 1D diffusion equations for temperature ($T$) and pore pressure ($p$) in direction normal to the fault \citep{noda2010three}:
\begin{equation}\label{EQ5}
    \frac{\partial T}{\partial t}=\alpha_\mathrm{th} \frac{\partial^2 T}{\partial y^2} + \frac{\tau V}{\rho cw \sqrt{2\pi}}\exp \left(\frac{-y^2}{2w^2}\right),
\end{equation}
\begin{equation}\label{EQ6}
    \frac{\partial p}{\partial t}=\alpha_\mathrm{hy} \frac{\partial^2 p}{\partial y^2} + \Lambda \frac{\partial T}{\partial t},
\end{equation}
where $\alpha_\mathrm{th}$ is the thermal diffusivity, $\rho c$ the volumetric heat capacity, $\tau$ the shear strength, $V$ the slip velocity, $w$ the shear-zone half-width, $\alpha_\mathrm{hy}$ the hydraulic diffusivity and $\Lambda$ the pore pressure increase per unit increase in temperature.
The 1D approximation is valid since dynamic rupture time scales are short compared to the time scales of diffusion along the fault \citep{zhu2020fault}.
We follow \cite{noda2010three}, \cite{noda2009earthquake} and \cite{andrews2002fault} in assuming that the shear strain rate has a Gaussian profile in the fault-normal direction, while \cite{bizzarri2006thermal_part1, bizzarri2006thermal_part2} assumed a uniform profile. 

We set $\alpha_\mathrm{hy}$ as constant in the velocity-weakening region and increasing by $\Delta\alpha_\mathrm{hy}$ in the velocity-strengthening region, with the following spatial distribution:
\begin{equation}\label{EQ7}
    \Delta \alpha_\mathrm{hy}(x,z)=\Delta \alpha_{hy0} \, [1-B_1(x;H,h) B_3(z;H,h)],
\end{equation}
where $\Delta \alpha_{hy0}=$ \SI{1}{\m^2\per\s} and
\begin{equation}\label{EQ8}
B_3(z;W,w)=
    \begin{cases}
    1, \,\, z \leq H
    \\
    0.5 \left[1+\tanh\left(\frac{h}{z-H-h}+\frac{h}{z-H} \right) \right], \,\, H<z<H+h
    \\
    0, \,\, z \geq H+h.
    \end{cases}
\end{equation} \\

\paragraph{}
Laboratory experiments and previous numerical simulations accounting for TP suggest that $\alpha_\mathrm{hy}$ from $10^{-5}$ to $10^{-2} \, \SI{}{m^2/s}$ and $w$ from $1$ to $100 \, \SI{}{mm}$ \citep{kranz1990hydraulic, andrews2002fault,  wibberley2002hydraulic, wibberley2005earthquake, noda2009earthquake, schmitt2015nucleation, noda2010three, perry2020nearly, lambert2021propagation}. Here, we vary the values of both parameters as indicated in Table \ref{tab:MP-FP-TP-IC}. 
The $\alpha_\mathrm{hy}$ and $w$ ranges in Table \ref{tab:MP-FP-TP-IC} allow rupture to propagate over entire fault as mostly subshear (only local supershear occurs) with smooth arrest providing realistic rupture process and preventing very high temperatures (T $>$ 1200 K).

\paragraph{}
We perform 49 dynamic rupture simulations considering 7 different values of $\alpha_\mathrm{hy}$ and $w$ and their combinations (Table \ref{tab:MP-FP-TP-IC}). The initial on-fault shear stresses and nucleation procedure are described in Appendix \ref{appendix:IniCond} and Appendix \ref{appendix:NucProc}, respectively, of the electronic supplement.

\subsection*{Numerical method and discretisation}
\paragraph{}
We use the open source software SeisSol, based on the Arbitrary high-order accurate DERivative Discontinuous Galerkin method (ADER-DG)\citep{dumbser2006arbitrary, pelties2014verification, uphoff2017extreme}, to solve the coupled dynamic rupture and 3D wave propagation problem including TP effects. We use a 4th-order accurate numerical scheme with basis functions of polynomial order 3. 
The fault is planar and of size $44 \times 22 \ \SI{}{km}$.
The computational domain is a box of size $220 \times 180 \times 40 \, \SI{}{km}$ centered on the fault.
It is discretized by an unstructured mesh of 13.6 million tetrahedral elements, 
with element edge lengths of 1500 m in most of the domain and refined to 200 m close to the fault. 
Our simulations resolve the seismic wavefield with frequencies up to 3.5 Hz close to fault and nearly 0.5 Hz in the more coarsely meshed regions. 
We store rupture parameters at 2000 randomly distributed on-fault receivers (Figure \ref{fig:rec}A) and ground velocity time series at 1516 surface receivers distributed along a set of Joyner-Boore distances ($R_{JB}$, Figure \ref{fig:rec}B). 
With these settings, simulating 55 seconds of rupture process and wave propagation requires around 3.6 hours on $256 \times 64$ Haswell cores of the Shaheen II supercomputer at KAUST.

\subsection*{A spectral implementation of thermal pressurization for 3D Discontinuous Galerkin dynamic rupture methods}

\paragraph{}
Implementations of TP for dynamic rupture modeling vary in efficiency and flexibility.
Explicit Finite Difference (FD) schemes are simple to implement and allow for heterogeneous properties, such as diffusivity, within the bulk. \citet{noda2009earthquake} discretized spatial derivatives on a 1D grid orthogonal to the fault using a central difference scheme, and time derivatives using a simple explicit Euler scheme. However, such FD methods require small time steps and large memory due to the interpolation between major time steps, which involves storing the slip history at all sub-time steps. The spectral method \citep{noda2010three} allows for larger time steps, has excellent error properties and is memory efficient. However, it is restricted to homogeneous TP parameters along the direction normal to the fault. The spectral implementation can exploit the symmetry of the diffusion equation for similar material parameters on both sides of the fault.

\paragraph{}
Here, we couple a semi-analytic, stable, precise and efficient spectral diffusion solver introduced by \citet{noda2010three} with 3D dynamic rupture modeling in the HPC-empowered ADER-DG software SeisSol using a truncated Fourier expansion to evaluate the second-order space derivatives. The approximation related to truncation is very accurate:  the heat source is assumed to have a Gaussian profile and the Fourier spectrum of a Gaussian decays rapidly.
We first pre-calculate the inverse Fourier coefficients using a trapezoidal rule. Next, we calculate the current shear stresses and slip  and the corresponding shear heat source in the spectral domain. For the next time-step, temperature and pressure are then updated in the wavenumber domain. Lastly, we recover temperature and pore pressure in the space domain by inverse Fourier transformation using the pre-calculated coefficients.

\paragraph{}
In contrast to finite-element (FE) methods, the basis functions here are non-zero over the whole domain leading to a global approach.
We implement the update scheme for this TP solver coupled to strongly rate-dependent friction dynamic rupture following \citet{kaneko2008spectral}. 
We observe lower computational overhead when comparing SeisSol's spectral TP implementation to an FE TP approach, and comparable computational overhead to an FD TP solver, in comparison to respective SeisSol, FEM or FD simulations without TP. Similar to \citet{noda2010three}, we assume that the heat source remains constant during a time-step, which we expect to limit accuracy. \citet{li2007numerical} show how to construct a fourth order accurate spectral method for the heat equation which may be an interesting avenue for future TP implementations in dynamic rupture solvers.

\subsection*{Verification of the 3D TP implementation in SeisSol}
\paragraph{}
While analytical solutions are available for a shear heating zone of zero thickness \citep{rice2006heating} or for a short-term constant heat input under constant slip rate \citep{noda2010three}, more complex 3D dynamic rupture simulations with TP have to be verified without available analytical references.
Here, we verify the 3D TP implementation in SeisSol by running the most recent Southern California Earthquake Center (SCEC) dynamic rupture community benchmark TPV105-3D \citep{harris2011verifying, harris2018suite, gabriel20203d}.
Conducting an earlier 2D TP benchmark (TPV105-2D from 2011) revealed important physical challenges, such as the occurrence of high temperatures implying wide-spread frictional melting, and numerical challenges, including lack of accuracy and large differences between results with different time integration schemes.

\begin{figure}[H]
    \centering
    \includegraphics[width=1.0\textwidth]{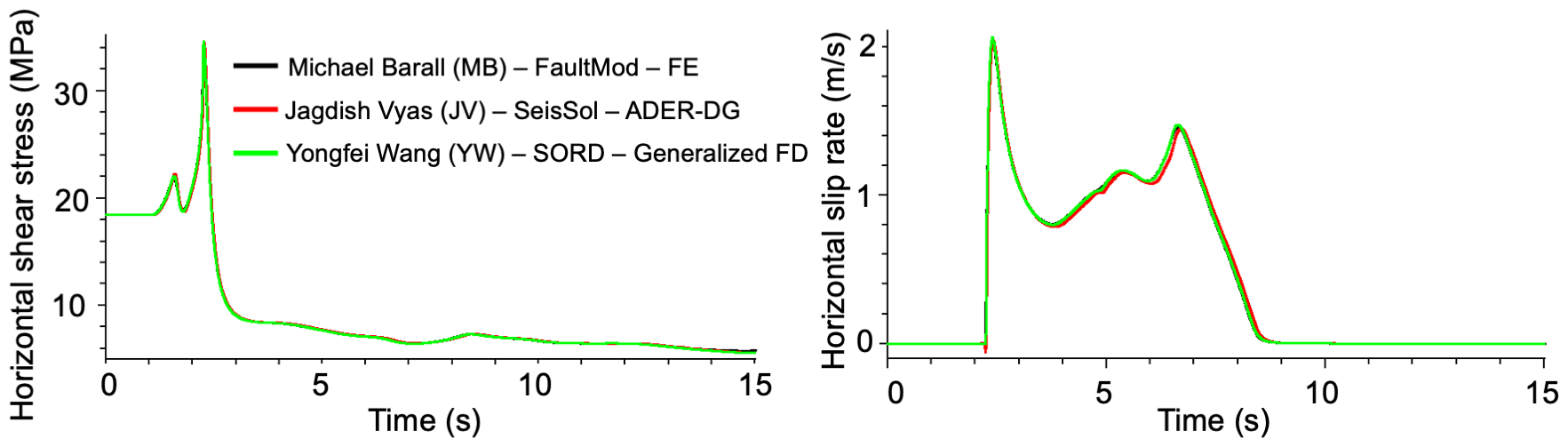}
    \caption{Comparison of the temporal evolution of horizontal shear stress and slip rate, computed using three different numerical methods as part of SCEC TPV105-3D benchmark exercise for validating dynamic rupture codes.}
    \label{fig:Hor-SR-Stress}
\end{figure}

\paragraph{}
We compare the SeisSol solution for the TPV105-3D benchmark exercise against other dynamic rupture codes. Figure \ref{fig:Hor-SR-Stress} shows a comparison for horizontal shear stress and slip rates computed using three different simulation methods as part of the SCEC TPV105-3D benchmark exercise. Three dynamic rupture codes accounting for TP (FaultMod, SeisSol and SORD) using different numerical schemes (FE, ADER-DG and generallized FD, respectively) for modeling rupture dynamics are compared. The SeisSol solution (JV) achieves excellent agreement with FaultMod (MB) and SORD (YW) (Figure \ref{fig:Hor-SR-Stress}). 

\paragraph{}
The 3D benchmark settings form the basis of the rupture simulations performed here. The benchmark setup \citep{ulrich_TPV1053d} identifies a relatively narrow (with respect to observational uncertainties) parameter range to realise realistic earthquake ruptures on a single, planar fault. It requires a velocity-strengthening transition layer at shallow depth to limit localised supershear rupture and spatially varying hydraulic diffusivity aligned with the rate-and-state parameters for smooth rupture stopping to aid realistic ground motion analysis. The chosen parameter values prevent unrealistically high temperatures (T$>$1200 K) that would imply melting. It features spatially heterogeneous TP characteristics, prevents high frictional energy at shallow depths and sets shear traction at 41\% of normal traction. The challenges in defining physically plausible 3D benchmark parameters due to the strong trade-offs of rupture nucleation, smooth rupture termination without reaching fault ends, rapid dynamic weakening effects, realistic heat production, and fault stress and strength initial conditions motivate the analysis in this study.

\section*{TP influence on rupture characteristics and ground shaking} 
\paragraph{}
We analyze the temporal evolution of selected rupture parameters and of ground velocity, and statistically analyse them to better understand TP effects on earthquake rupture and ground motion.  

\subsection*{Rupture parameter evolution and statistics}
\begin{figure}[H]
    \centering
    \includegraphics[width=1.0\textwidth]{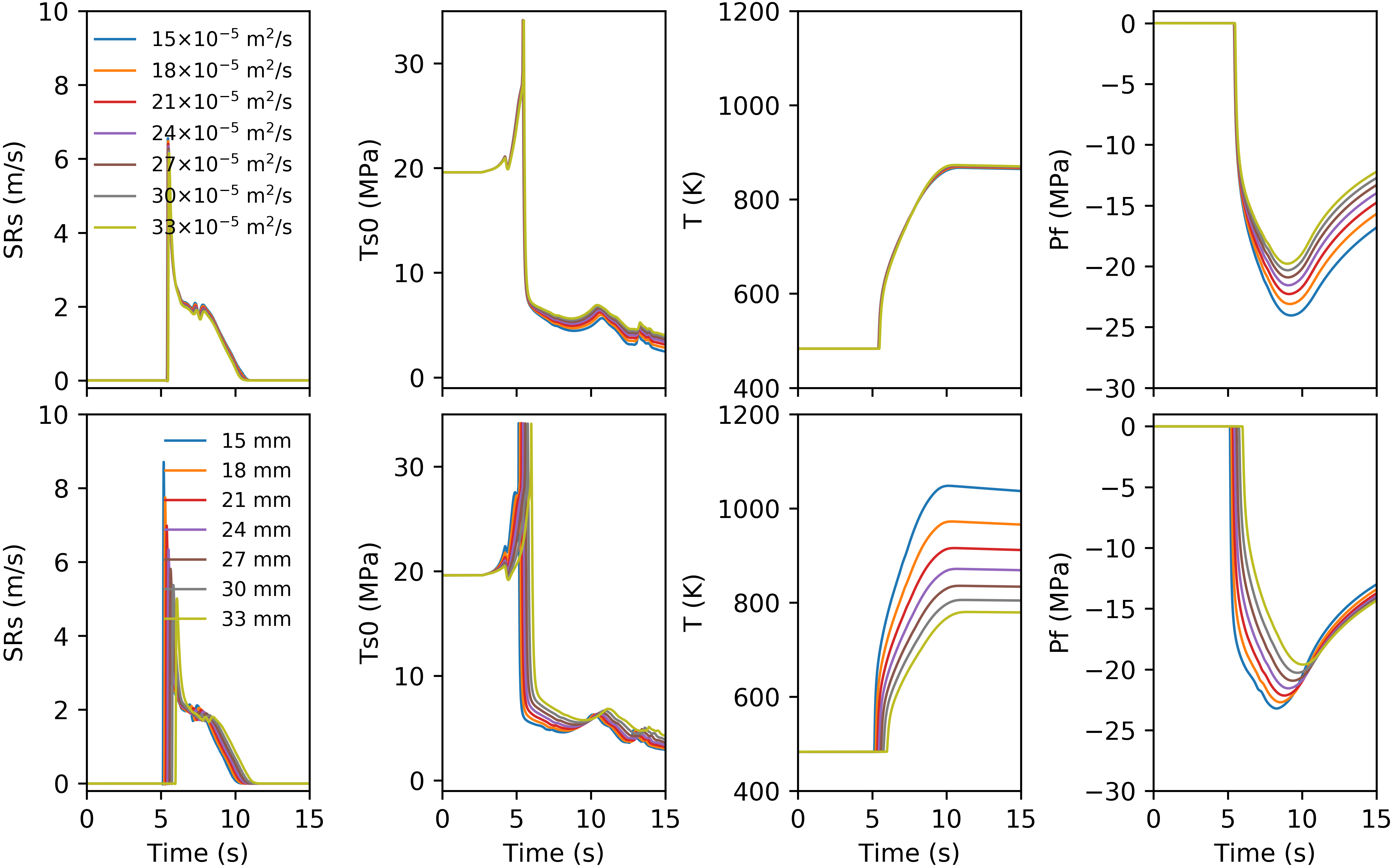}
    \caption{Temporal evolution of along strike slip rate (SRs), shear traction along strike (Ts0), temperature (T) and fluid pressure (Pf) at receiver s2 (Figure \ref{fig:rec}) for varying hydraulic diffusivity $\alpha_\mathrm{hy}$ (top row) and shear-zone half-width $w$ (bottom row).}
    \label{fig:s2-SRs}
\end{figure}

\paragraph{}
We analyze the temporal evolution of rupture parameters as a function of TP parameters. Figure \ref{fig:s2-SRs} shows the effects of varying hydraulic diffusivity $\alpha_\mathrm{hy}$ (in $m^2/s$) and shear-zone half-width $w$ (in $mm$) on the temporal evolution of four rupture variables: along-strike slip rate SRs, total traction Ts0, temperature T and pore pressure Pf at location s2 on the fault indicated on Figure \ref{fig:rec}A. We observe a stronger impact of $w$ on SRs, Ts0, T and Pf compared to $\alpha_\mathrm{hy}$. The onset time of 
rupture is delayed as we increase $w$ whereas rupture speed remains nearly unaffected when varying $\alpha_\mathrm{hy}$. Similarly, the absolute peak values of slip rate, temperature and pore pressure decrease more pronouncedly when increasing $w$ than when increasing $\alpha_\mathrm{hy}$. The rupture evolution at two other locations on the fault (s1 and s3) shows similar characteristics as observed at location s2 (compare electronic supplement Figures \ref{fig:s1-SRs} and \ref{fig:s3-SRs} with Figure \ref{fig:s2-SRs}). 

\begin{figure}[H]
    \centering
    \includegraphics[width=1.0\textwidth]{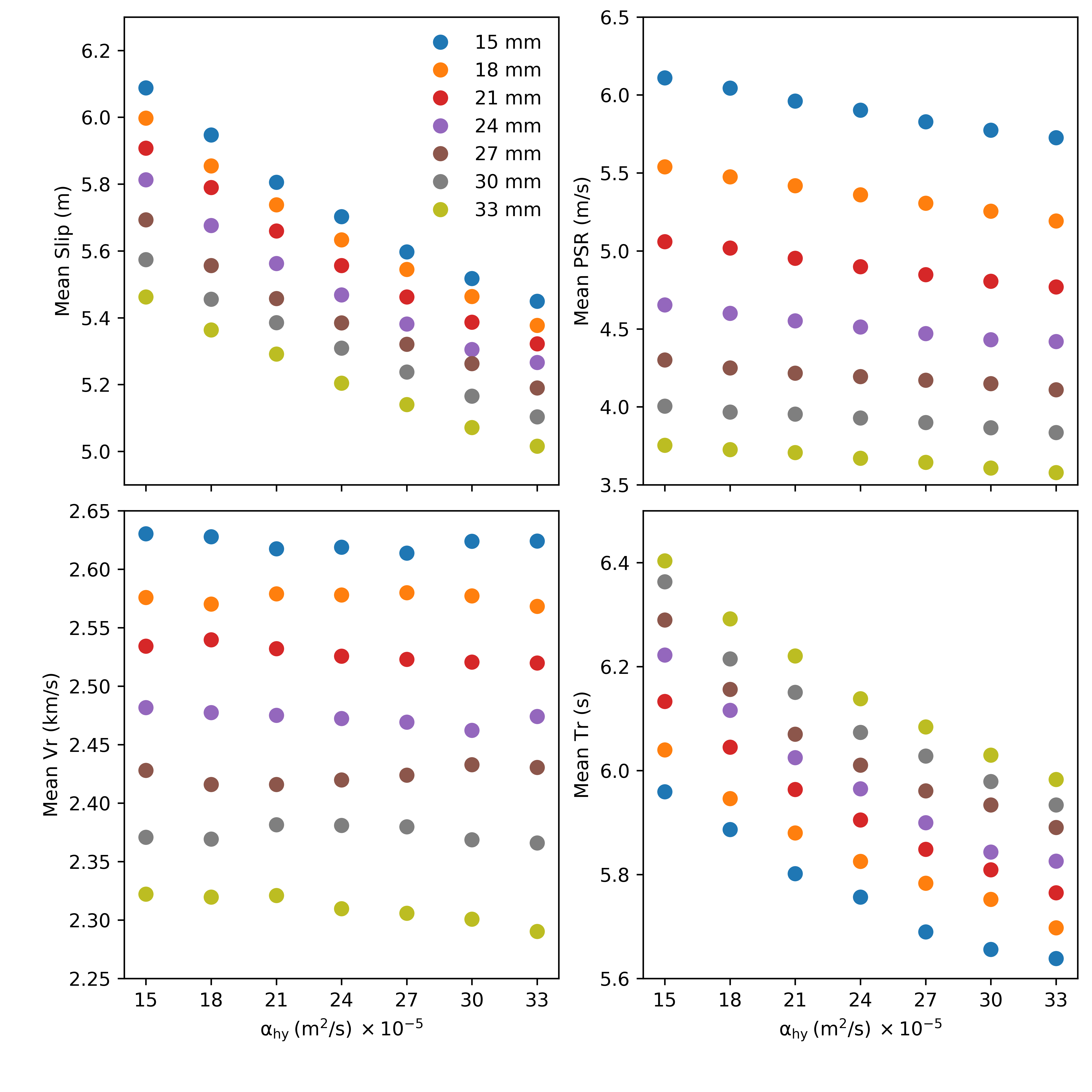}
    \caption{Mean slip, peak slip rate (PSR), rupture speed (Vr) and rise time (Tr) variations with respect to hydraulic diffusivity $\alpha_\mathrm{hy}$ for given shear-zone half-width $w$ (color-coded).}
    \label{fig:mean-slip-wrt-alpha}
\end{figure}

\paragraph{}
We quantify TP effects on earthquake source processes by statistically analyzing kinematic rupture properties across all 2000 on-fault receivers. 
We first analyse the mean values of slip, PSR, Vr and Tr. We compute Vr as $1/\| \nabla(\mathrm{Tr}) \|$, using 2D basis functions across each dynamic rupture element face. 
Figure \ref{fig:mean-slip-wrt-alpha} shows variations in mean slip, PSR, Vr and Tr for varying $\alpha_\mathrm{hy}$ (x-axis) for a given $w$ (color-coded). Mean slip and Tr decrease by nearly 10\% and 6\%, respectively, as we double the value of $\alpha_\mathrm{hy}$ for a given $w$, whereas PSR and Vr remain nearly constant. 
As we increase $\alpha_\mathrm{hy}$ the fluid diffuses faster. This leads to an earlier reduction of pore pressure, increasing effective normal stress and therefore strengthening the fault. This explains the observed reduced fault slip with increasing $\alpha_\mathrm{hy}$. The fault slip reduction translates into a drop of earthquake magnitude Mw of almost 5\% when doubling $\alpha_\mathrm{hy}$ (supplementary Figure \ref{fig:Mw-var}). 

\begin{figure}[H]
    \centering
    \includegraphics[width=1.0\textwidth]{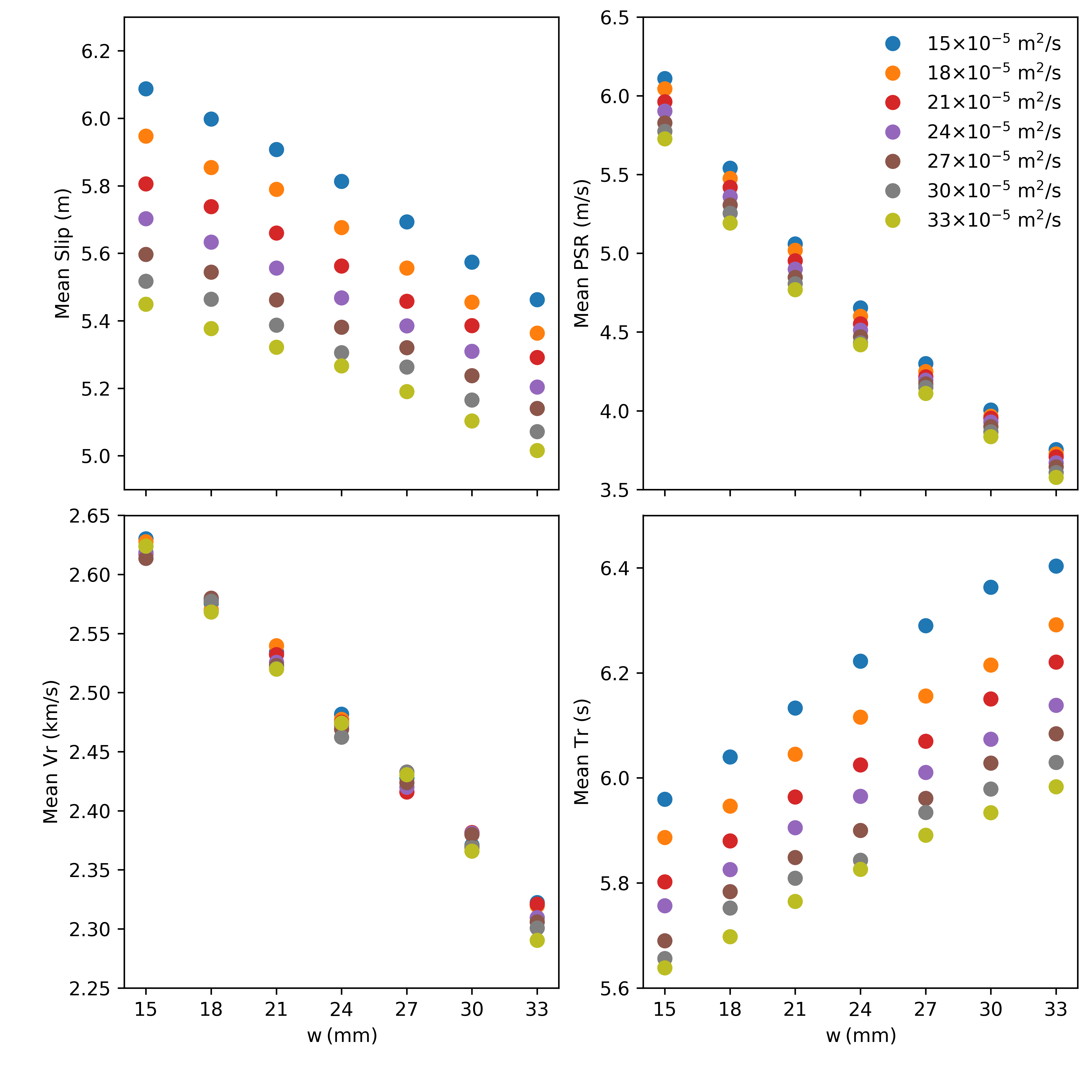}
    \caption{Mean slip, peak slip rate (PSR), rupture speed (Vr) and rise time (Tr) variations with respect to shear-zone half-width $w$ for given hydraulic diffusivity $\alpha_\mathrm{hy}$ (color-coded).}
    \label{fig:mean-slip-wrt-w}
\end{figure}

\paragraph{}
We analyze and compare effects of $w$ against $\alpha_\mathrm{hy}$ on kinematic rupture properties.  Figure \ref{fig:mean-slip-wrt-w} shows variations in mean slip, PSR, Vr and Tr for varying $w$ and fixed $\alpha_\mathrm{hy}$. We observe that mean slip, PSR and Vr decrease with increasing $w$ whereas mean Tr increases. The mean slip decreases by nearly 10\% whereas the mean Tr increases by almost 6\% as we double $w$, similar to the magnitude of changes found by varying $\alpha_\mathrm{hy}$. Hence, both $\alpha_\mathrm{hy}$ and $w$ almost equally influence mean slip and earthquake magnitude (supplementary Figure \ref{fig:Mw-var}). Both parameters have similar impact on mean Tr with opposite trends (decreasing and increasing, respectively). The mean PSR and Vr decrease by nearly 60\% and 14\%, respectively, as we double $w$, but are nearly unaffected by variations of $\alpha_\mathrm{hy}$ in the range considered (Figure \ref{fig:mean-slip-wrt-alpha}). Therefore, $w$ has a stronger impact than $\alpha_\mathrm{hy}$ on the overall earthquake rupture process, especially on PSR, which may also strongly affect ground shaking.

\subsection*{Correlation of kinematic rupture parameters}
\paragraph{}
We now analyze correlations between pairs of rupture parameters. Specifically, we calculate the Pearson correlation coefficient, $r$, between six pairs of source parameters: (slip, PSR), (slip, Vr), (slip, Tr), (PSR, Vr), (PSR, Tr) and (Vr, Tr). The linear correlation coefficient $r$ does not completely capture potential non-linear correlations, but it is the metric considered by \citet{schmedes2010correlation}, which allows us to compare our simulation results with theirs.
As in \cite{schmedes2010correlation}, we exclude on-fault receivers around the rupture nucleation and arrest areas, by considering only receivers contained in the red dashed rectangle of Figure \ref{fig:rec}.

\begin{figure}[H]
    \centering
    \includegraphics[width=1.0\textwidth]{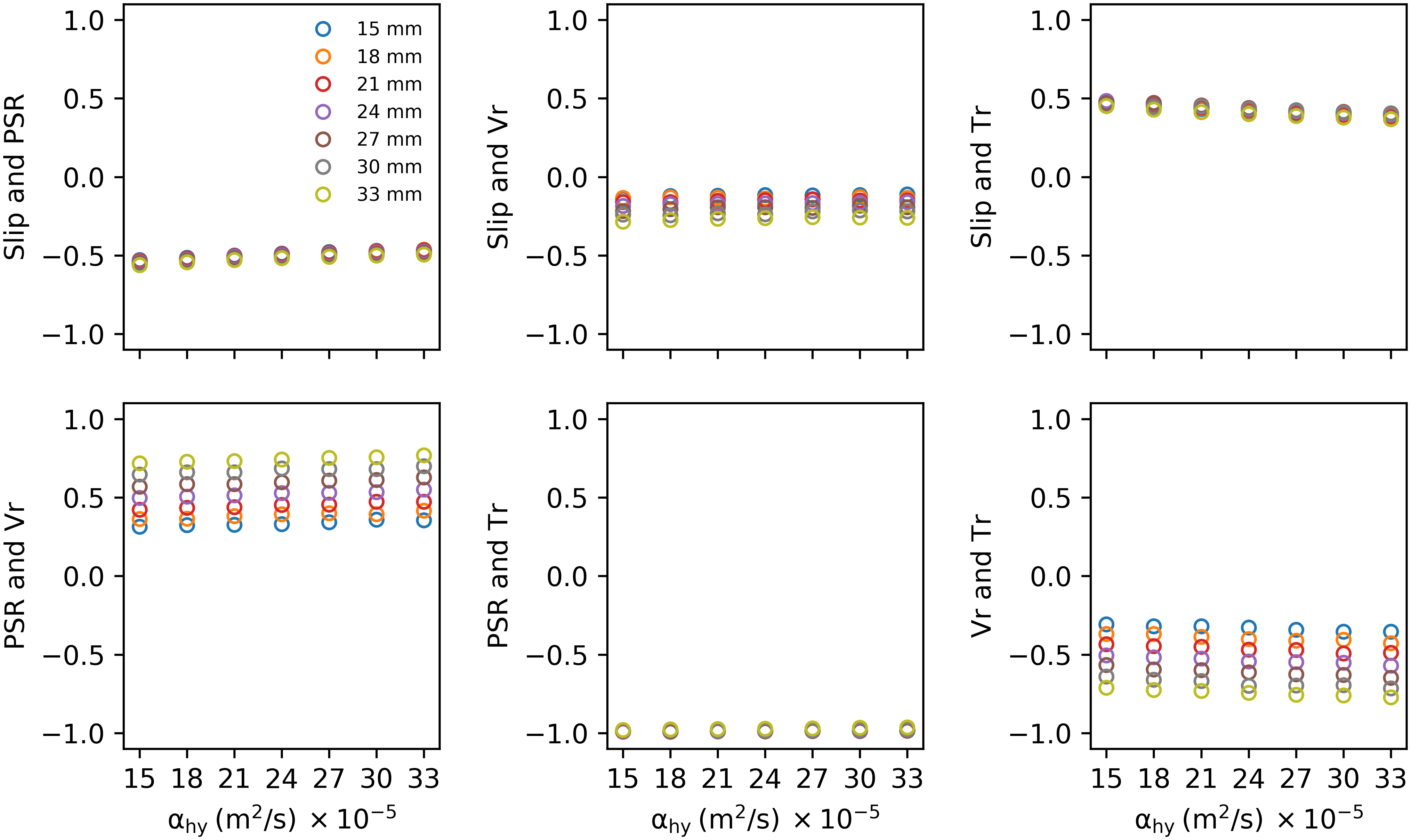}
    \caption{Correlations between rupture parameters as a function of $\alpha_\mathrm{hy}$ for given $w$ (color-coded).  Correlations are based on the set of receivers contained in the red dashed rectangle in Figure \ref{fig:rec}.}
    \label{fig:rup-par-corr-wrt-alpha}
\end{figure}

\paragraph{}
We analyze correlations between kinematic rupture parameters as a function of $\alpha_\mathrm{hy}$. Figure \ref{fig:rup-par-corr-wrt-alpha} shows correlation among rupture parameters with varying hydraulic diffusivity $\alpha_\mathrm{hy}$ for given shear-zone half-width $w$. The correlations between the six selected source parameter pairs are nearly insensitive to changes in $\alpha_\mathrm{hy}$. The (slip, PSR) and (Vr, Tr) are negatively correlated ($r \sim -0.5$), and (PSR, Tr) are strongly negatively correlated ($r \sim -1.0$). The (slip, Tr) and (PSR, Vr) are positively correlated ($r \sim 0.5$), whereas (slip, Vr) are almost uncorrelated ($r \sim -0.2$).

\begin{figure}[H]
    \centering
    \includegraphics[width=1.0\textwidth]{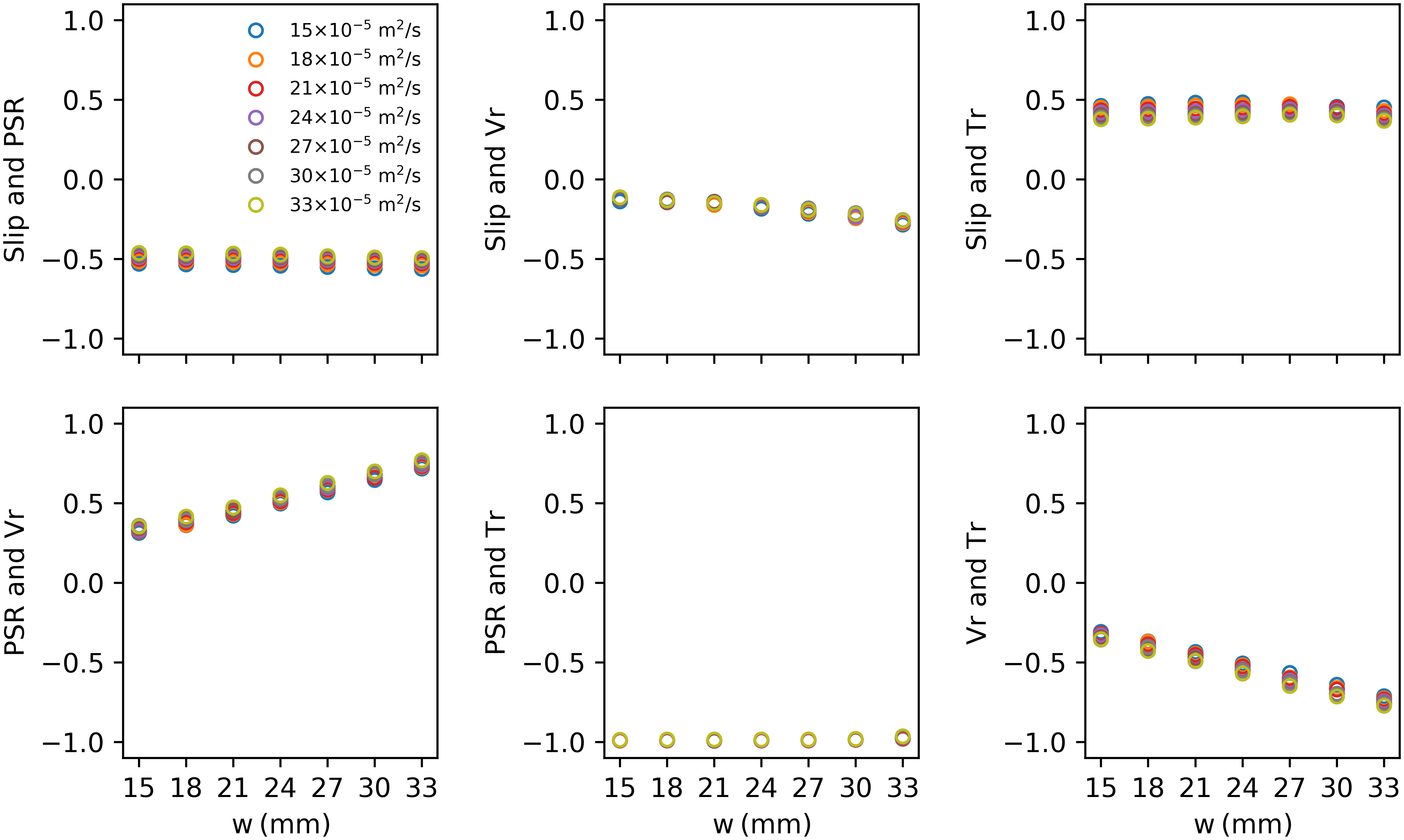}
    \caption{Correlations between rupture parameters as a function of $w$ for given color-coded $\alpha_\mathrm{hy}$. Correlations are based on the set of receivers contained in the red dashed rectangle in Figure \ref{fig:rec}.}
    \label{fig:rup-par-corr-wrt-w}
\end{figure}

\paragraph{}
We analyze and compare effects of $w$ against $\alpha_\mathrm{hy}$ on the correlations between kinematic rupture parameters. Figure \ref{fig:rup-par-corr-wrt-w} shows correlations among rupture parameters with varying $w$ for fixed $\alpha_\mathrm{hy}$. Correlations among (slip, PSR), (slip, Tr) and (PSR, Tr) are nearly insensitive to changes in $w$, which is similar to what we observed in Figure \ref{fig:rup-par-corr-wrt-alpha} when varying $\alpha_\mathrm{hy}$. However, (PSR, Vr) correlation increases from  $0.3$ to almost $0.7$ and (Vr, Tr) anticorrelation increases from $0.3$ to almost $0.7$ as we double $w$. The contribution of TP to dynamic weakening is substantial for the smallest explored $w$. Indeed, TP effects can strongly decrease the correlation among (PSR, Vr) and (Vr, Tr). We also find a slight increase in anticorrelation from $0.1$ to $0.3$ for (slip, Vr) when doubling $w$. All parameter pairs whose correlation increases with increasing $w$ involve rupture speed Vr. This suggests that $w$ may primarily affect Vr, which in turn may affect other kinematic rupture parameters such as slip, PSR and Tr.

\begin{figure}[H]
    \centering
    \includegraphics[width=1.0\textwidth]{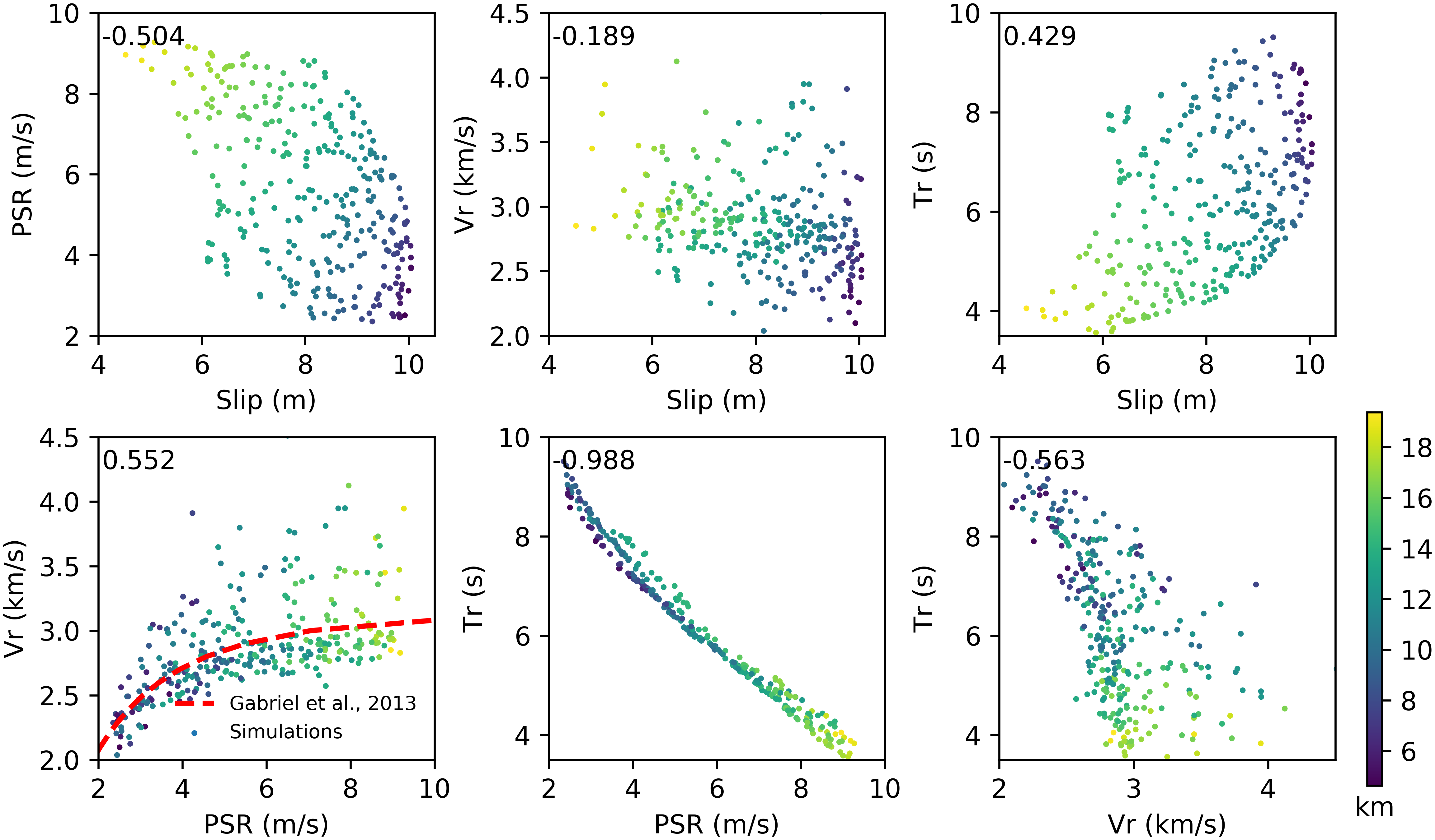}
    \caption{Correlations between rupture parameters, averaged over all 49 simulations, for six pairs of rupture variables and at all fault receivers. The correlation coefficient computed from the averaged dataset is shown in the top-left corner of each subplot. Dots are color-coded by hypocenter-receiver distance.}
    \label{fig:rup-par-corr-avg-Nsim}
\end{figure}

\paragraph{}
We compute the mean value of each rupture parameter across the 49 explored models at each on-fault receiver to understand the overall correlations among slip, PSR, Vr and Tr. 
Figure \ref{fig:rup-par-corr-avg-Nsim} shows these average correlations for six pairs of rupture variables. We observe that slip and PSR are negatively correlated ($r \sim -0.5$), which is in contrast to the positive correlation reported by \citet{schmedes2010correlation}. The low correlation we observe between slip and Vr is consistent with findings of \citet{schmedes2010correlation}. We find positive correlation for (slip, Tr; $r \sim 0.4$) in agreement with a positive relationship reported by \citet{schmedes2010correlation} with $r \sim 0.6$ to $0.7$. We find that PSR and Vr are positively correlated consistent with \citet{schmedes2010correlation}. 
We find negative correlation for (PSR, Tr; $r \sim -1.0$) and (Vr, Tr; $r \sim -0.6$) consistent with findings of negative relationship reported by \cite{schmedes2010correlation} ($r \sim -0.2$ for two pairs). Also, we show that theoretical estimates of (PSR, Vr) derived in \citet{gabriel2013source} match our simulation results (Figure \ref{fig:rup-par-corr-avg-Nsim}). The slightly different scaling factor between PSR and Vr estimated from our models (0.8) compared to the theoretical estimates from \cite{gabriel2013source} (0.65) can be explained by the use of 2D models without TP in \cite{gabriel2013source}.

\paragraph{}
The differences observed in the correlation among rupture parameters from our study with respect to \cite{schmedes2010correlation} can be partly attributed to dissimilar features of their rupture models, including the initial shear stress, friction law and inclusion of TP effects. We use depth dependent initial shear stress (see Appendix A of electronic supplement) whereas \cite{schmedes2010correlation} assume a spatially heterogeneous initial shear stress with power-law spectral decay. \cite{schmedes2010correlation} employ linear slip-weakening friction whereas we use rate-and-state friction with strong velocity-weakening, consistent with laboratory observations. Our simulations account for TP whereas \cite{schmedes2010correlation} did not.

\subsection*{Ground velocity comparison and PGV statistics}
\begin{figure}[H]
    \centering
    \includegraphics[width=1.0\textwidth]{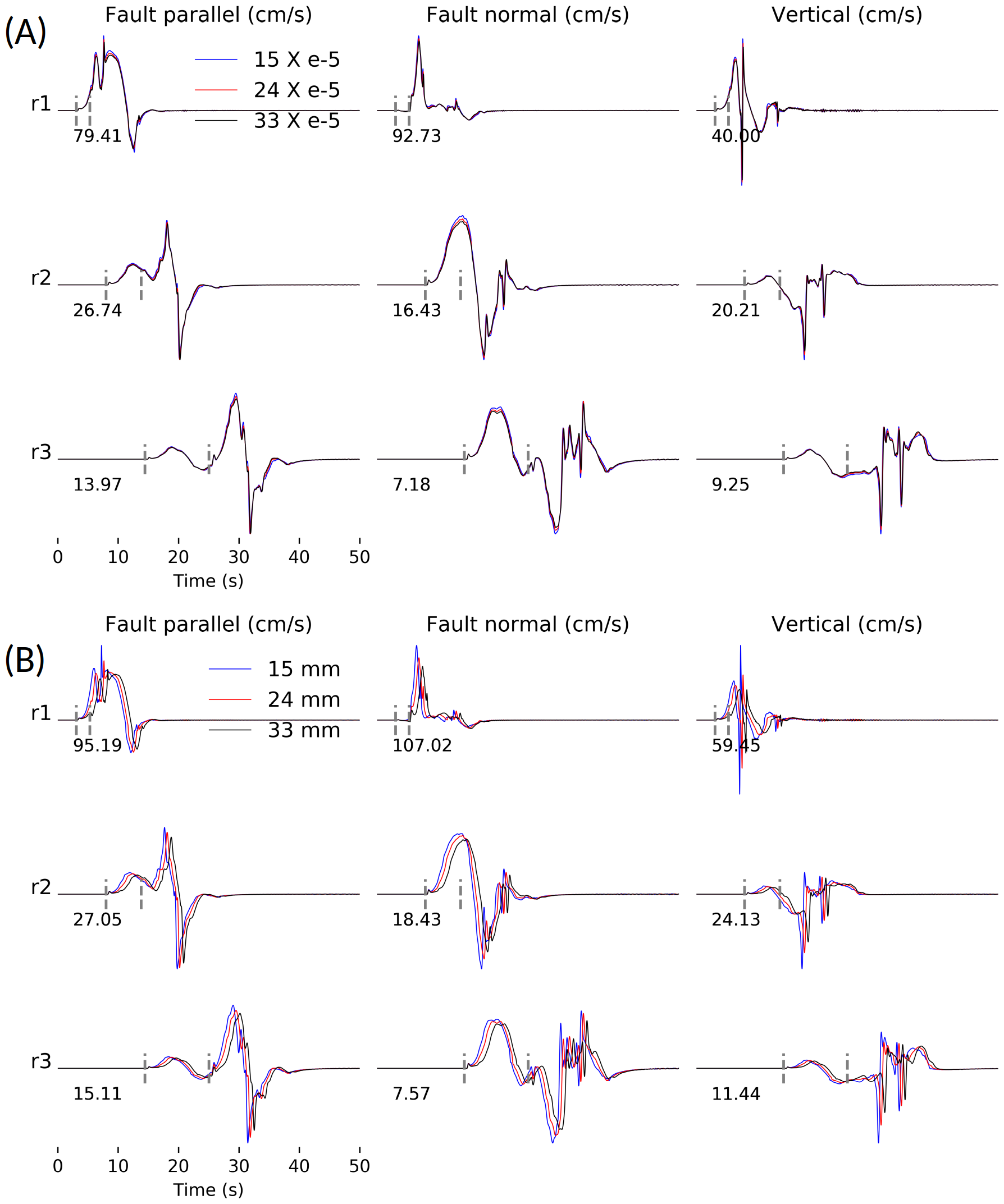}
    \caption{Comparison of ground velocity time histories (cm/s) at three receivers (r1–r3, Figure \ref{fig:rec}B) for varying (A) hydraulic diffusivity $\alpha_\mathrm{hy}$ and (B) shear-zone half-width $w$. Gray dashed bars show theoretical P- and S-wave arrival times from the hypocenter. Waveforms are normalized with respect to peak absolute value (indicated in left corner) for each component.
    }
    \label{fig:vel-comp}
\end{figure}

\paragraph{}
We analyze ground velocity time series as a function of TP parameters. Figure \ref{fig:vel-comp} compares three-component ground velocity waveforms for three values of $\alpha_\mathrm{hy}$ (15, 24, 33 X $10^{-5}$ $m^2/s$) and $w$ (15, 24, 33 mm) at receivers r1 to r3 (Figure \ref{fig:rec}B). Velocity waveforms are nearly unaffected by changes in $\alpha_\mathrm{hy}$ for all components and receivers. On the other hand, increasing $w$ delays the first arrival and reduces peak amplitude, while the waveform shape remains nearly similar (see, Figure \ref{fig:vel-comp}B).
Hence, $w$ seems to have a larger impact on ground velocity than $\alpha_\mathrm{hy}$.  

\begin{figure}[H]
    \centering
    \includegraphics[width=1.0\textwidth]{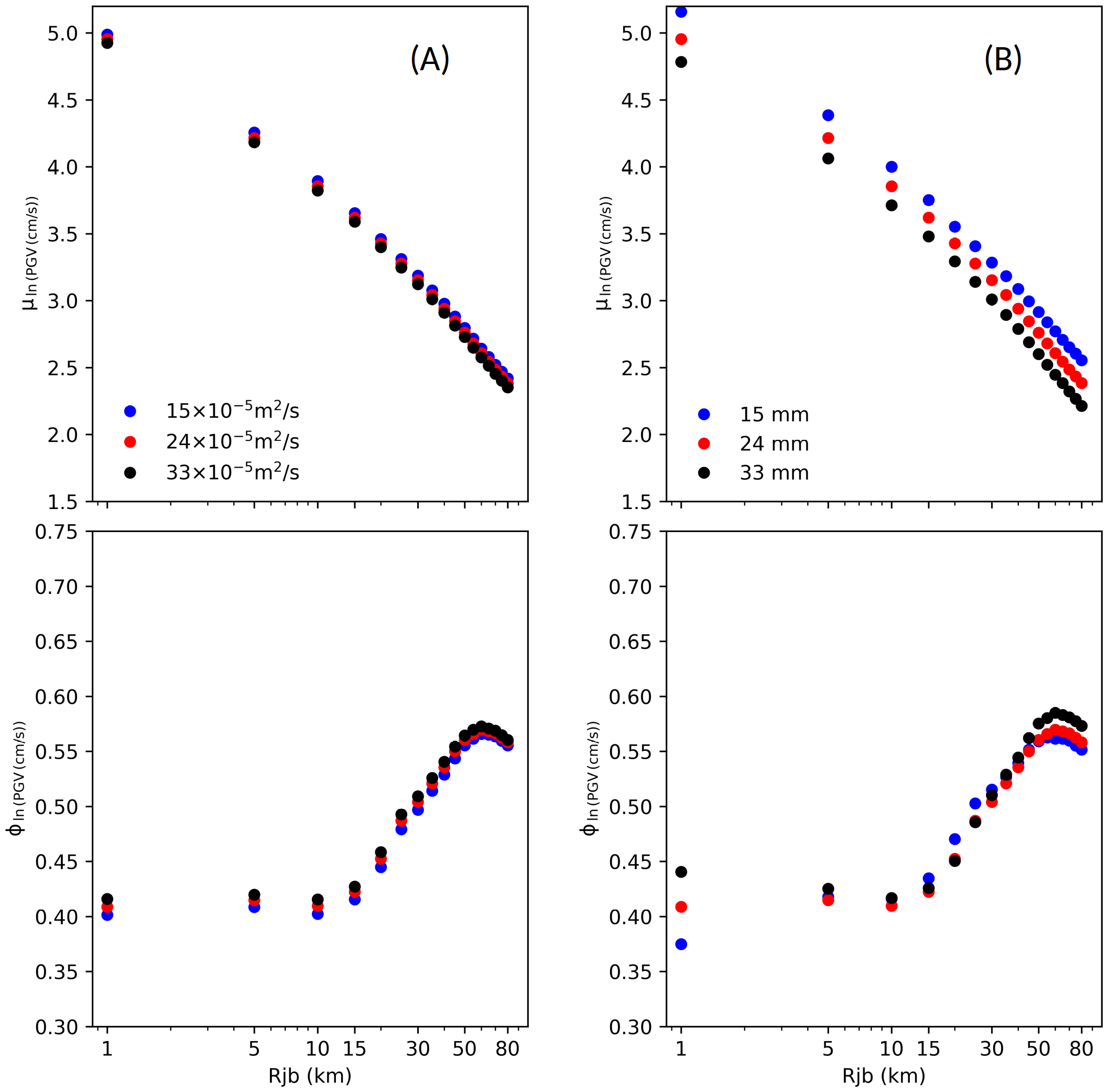}
    \caption{Distance dependence of mean $\mu_{\ln(\mathrm{PGV})}$ and standard deviation $\phi_{\ln(\mathrm{PGV}}$ of $\ln(\mathrm{PGV})$ for (A) varying hydraulic diffusivity $\alpha_\mathrm{hy}$ and (B) shear-zone half-width $w$. Colors represent  different $\alpha_\mathrm{hy}$ and $w$ values.
    }
    \label{fig:PGV-stat}
\end{figure}

\paragraph{}
We examine the mean and standard deviation of PGV across all receivers at a given $R_{JB}$ distance. We calculate the orientation-independent PGV metric GMRotD50 \citep{boore2006orientation}. Figure \ref{fig:PGV-stat} shows the distance dependence of the mean ($\mu_{\ln(PGV)}$) and the standard deviation ($\phi_{\ln(PGV)}$) of $\ln(PGV)$ for varying $\alpha_\mathrm{hy}$ and $w$. $\mu_{ln(PGV)}$ decreases much more with increasing $w$ than with increasing $\alpha_\mathrm{hy}$. The mean PGV decreases by nearly 15\% at 80 km $R_{JB}$ distance when $w$ is doubled, whereas it decrease by only 2\% when $\alpha_\mathrm{hy}$ is doubled. The variability $\phi_{\ln(PGV)}$ has a similar trend at all $R_{JB}$ distances for both $\alpha_\mathrm{hy}$ and $w$ except near the fault ($R_{JB}$ = 1 km). We also observe an increase in ground motion variability $\phi_{\ln(PGV)}$ beyond 15 km $R_{JB}$ which is consistent with the reporting of \cite{imtiaz2015ground} of increasing variability with distance for bilateral ruptures, and is therefore not specific to rupture accounting for TP. In summary, we observe that $w$ has a larger impact on ground velocity and PGV than $\alpha_\mathrm{hy}$.

\section*{Discussion and conclusions} 
\paragraph{}
We investigate the role of two key TP parameters, hydraulic diffusivity $\alpha_\mathrm{hy}$ and shear-zone half-width $w$, on earthquake rupture dynamics, associated kinematic source properties and ground shaking. 
We conduct 3D dynamic rupture simulations considering a rate-and-state dependent friction law with strong velocity-weakening, coupled to TP.

\paragraph{}
We find that $w$ has a stronger effect on rupture dynamics, kinematic rupture properties and ground shaking than $\alpha_\mathrm{hy}$.
Mean slip and rise time decrease with increasing $\alpha_\mathrm{hy}$, whereas mean peak slip-rate and rupture speed remain nearly constant. Mean slip, peak slip-rate and rupture speed decrease with increasing $w$, whereas mean rise time increases. Also, $w$ distinctly affects the correlation between rupture parameters, especially between parameter pairs (slip, rupture speed), (peak slip-rate, rupture speed), (rupture speed, rise time), whereas $\alpha_\mathrm{hy}$ has a negligible effect on rupture-parameter correlations. Mean PGV decreases faster with increasing $w$ than with increasing $\alpha_\mathrm{hy}$, whereas ground motion variability has similar trends. 

\begin{figure}[H]
    \centering
    \includegraphics[width=1.0\textwidth]{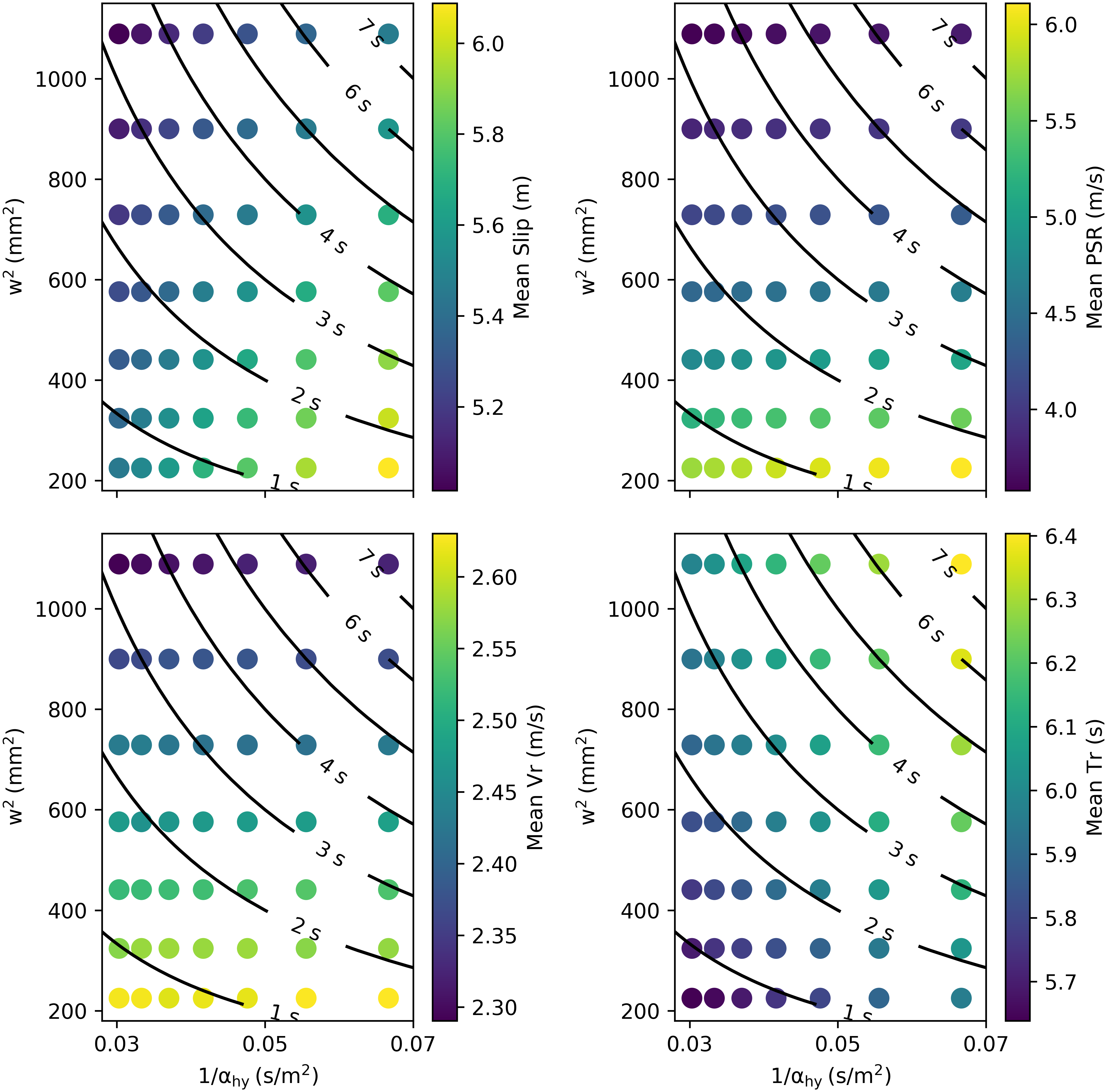}+
    \caption{Variations of mean slip, peak slip rate, rupture speed and rise time as function of 1/$\alpha_\mathrm{hy}$ and $w^2$. Contour lines mark diffusion time (eq. \ref{EQ10})}
    \label{fig:rupPar-alpha-w-t_diff}
\end{figure}

\paragraph{}
The dominant impact of $w$ compared to $\alpha_\mathrm{hy}$ on rupture dynamics can be interpreted by considering the characteristic time $t_\mathrm{diff}$ of the diffusion process:
\begin{equation}\label{EQ9}
    t_\mathrm{diff} \propto \frac{w^2}{\sqrt{ \alpha_\mathrm{th}^2 + \alpha_\mathrm{hy}^2 }}.
\end{equation}
Since $\alpha_\mathrm{th} << \alpha_\mathrm{hy}$, 
\begin{equation}\label{EQ10}
    t_\mathrm{diff} \approx \frac{w^2}{\alpha_\mathrm{hy}}.
\end{equation}
%
In particular, $t_\mathrm{diff}$ increases by nearly four times as we double $w$, whereas it decreases by only two times as we double $\alpha_\mathrm{hy}$. Figure \ref{fig:rupPar-alpha-w-t_diff} shows variations in kinematic rupture parameters (slip, PSR, Vr and Tr) as a function of $\frac{1}{\alpha_\mathrm{hy}}$ and $w^2$, along with contours of diffusion time. We find that rupture speed is inversely correlated to $t_\mathrm{diff}$, whereas rise time is positively correlated to $t_\mathrm{diff}$. Slip and PSR have more complex relations with $t_\mathrm{diff}$. The variations in PSR are larger for varying $w$ compared to $\alpha_\mathrm{hy}$, which is consistent with the variation of $t_\mathrm{diff}$ as a function of $w$ and $\alpha_\mathrm{hy}$. 

\paragraph{}
Analytical scaling relations have been developed for 2D steady-state pulse-like ruptures under TP \citep{garagash2012seismic}. Their Fig. 5a shows that pulse rise-time scales with the characteristic diffusion time across the shear zone for a given value of prestress: fault re-strengthening due to pore pressure diffusion away from the shear zone becomes effective at this time scale.
Similar effects may qualitatively explain our findings of strong influence of shear-zone half-width $w$ and the weaker influence of hydraulic diffusivity $\alpha_\mathrm{hy}$ on 3D heterogenous rupture pulse characteristics such as average rise time. 
Our modeled rupture pulses have relatively long rise times (on the order of 5 s), and thus healing will not only be governed by pore pressure diffusion as in the 2D steady-state pulse but also by geometrical effects such as the limited width of the seismogenic zone \citep{weng2019dynamics}.

\paragraph{} 
Our study considers constant values of hydraulic diffusivity and shear-zone half-width over a large portion of the fault (the velocity-weakening portion of the fault in Figure \ref{fig:rec}A), whereas complex spatial variations of TP parameters are expected in nature. Our simulations suggest that the effects on kinematic parameters and ground-motions decrease, especially for PSR (by nearly 60\%) and  for mean PGV (by nearly 15\%) respectively, as we double $w$. We do not consider path effects, such as seismic wavefield scattering and intrinsic attenuation, as our objective is to understand source effects due to TP. Therefore, future extension of this research could include other physical processes such as realistic high-frequency radiation due to fault roughness, rupture complexity due to heterogeneous initial shear stress, seismic scattering due to small scale heterogeneities and plastic deformation of material, to gain understanding on how TP combined with these processes affects rupture dynamics and ground-motion at high frequencies \citep{imperatori2012sensitivity, shi2013rupture, mai2018accounting, wollherr2018off, vyas2018mach, vyas2021characterizing}. The combined effects of several physical processes can reduce (or enhance) TP effects on rupture dynamics making it a second-order (or primary) process governing ground-shaking and PSHA. 

\paragraph{}
In summary, our physics-based dynamic rupture simulations accounting for TP develop an understanding on how hydraulic diffusivity and shear-zone half-width affects rupture dynamics, associated kinematic properties and ground-shaking. Our findings on how variations in $\alpha_\mathrm{hy}$ and $w$ affect slip, PSR, Vr, Tr and PGV provide useful insight for the earthquake source modeling community and hazard assessment community about rupture properties and ground-motion behaviour from earthquakes in fluid-rich environments such as geo-reservoirs and subduction zones where TP can be a dominant dynamic weakening mechanism. \\~\\~\\

\section*{Data and Resources} 
\paragraph{}
We use the open-source software SeisSol (\url{https://github.com/SeisSol/SeisSol}, git commit hash: 2f4942e8de21aebd56e179eb11837acd40ae4c07, last accessed October, 2022) to simulate earthquake rupture process and ground-motions. All SeisSol input files required to reproduce our simulations will be made available as a zenodo repository upon publication. We generated nearly 2.0 TB data to analyze the effects of pore fluid thermal pressurization on rupture dynamics and ground-shaking which can be provided via personal communication upon request. This article comprises electronic supplemental material on temporal evolution of slip rate, total traction, temperature and pressure at different on-fault locations. It also describes effects of hydraulic diffusivity and shear-zone half-width variations on moment magnitude. The details of initial shear stress and rupture nucleation method are also provided in the electronic supplement. The 3D benchmark exercise by \citet{ulrich_TPV1053d} is available on SCEC website (\url{https://strike.scec.org/cvws/download/TPV105-3D-description.pdf}, last accessed October, 2022).

\section*{Acknowledgments} 
\paragraph{}
We acknowledge the work and helpful discussions of and with Stephanie Wollherr, Sebastian Anger and Kadek Palgunadi regarding the implementation of thermal pressurisation in SeisSol. We also thank Michael Barall and Yongfei Wang for uploading their solutions to SECEC TPV105-3D benchmark exercise platform to validate their dynamic rupture codes. The research presented in this paper is supported by King Abdullah University of Science and Technology (KAUST) in Thuwal, Saudi Arabia, under grants BAS/1/1339-01-01 and URF/1/3389-01-01 (JCV and PMM).
AAG and TU acknowledge the support of the European Research Council (ERC) under the European Union’s Horizon 2020 research and innovation programme (ERC Starting Grant TEAR, Grant agreement No. 852992 and Center of Excellence ChEESE, Grant agreement No. 823844), the Southern California Earthquake Center (SCEC Grant No. 22135) and the National Science Foundation (NSF Grant No. EAR-2121666). 
Earthquake rupture dynamics and ground-motion modeling have been carried out using the KAUST Supercomputing Laboratory (KSL), and we acknowledge the support of the KSL staff. We also thank Kadek Palgunadi and SeisSol software team for their help regarding software installation.

\newpage
\bibliographystyle{apalike}
\bibliography{Ref}

\newpage
\renewcommand{\listtablename}{List of table caption}
{\renewcommand*\numberline[1]{Table \,#1:\space}
\listoftables
}
\renewcommand{\listfigurename}{List of figure captions}
{\renewcommand*\numberline[1]{Figure \,#1:\space}
\listoffigures
}

\newpage
\section*{Author's full physical mailing addresses}
\paragraph{}
Dr. Jagdish Chandra Vyas,\\
Building 1, Level 3, office 3111,\\
King Abdullah University of Science and Technology (KAUST),\\
Jeddah, Thuwal, 23955, Saudi Arabia.\\

\paragraph{}
Prof. Alice-Agnes Gabriel,\\
Department of Earth and Environmental Sciences, Room: C 416,\\
\normalsize{Ludwig-Maximilians-Universit\"at M\"unchen,}\\
Theresienstr 41,\\
80333 Munich, Germany.\\

\paragraph{}
Dr. Thomas Ulrich,\\
Department of Earth and Environmental Sciences, Room:  C 445,\\
\normalsize{Ludwig-Maximilians-Universit\"at M\"unchen,}\\
Theresienstr 41,\\
80333 Munich, Germany.

\paragraph{}
Prof. Paul Martin Mai,\\
Building 1, Level 3, office 3114,\\
King Abdullah University of Science and Technology (KAUST),\\
Jeddah, Thuwal, 23955, Saudi Arabia.\\

\paragraph{}
Jean Paul Ampuero, \\
Geoazur Laboratory, Université Côte d'Azur, \\
250 rue Albert Einstein, Sophia Antipolis, 06560 Valbonne, France.

\newpage 
\section*{List of table}
\begin{table}[H]
    \centering
    \setstretch{1.5}
    
    \caption{Modeling parameters and physical characteristics}
    \vspace{0.5em}
    
    \resizebox{\textwidth}{!}{
        \begin{tabular}{*{4}{c}}

        \hline
        & Variable & Notation & Value 
        \\
        \hline
        Material Properties &  &  & \\
        & P-wave velocity  & $c_p$ & $\SI{6}{km/s}$ \\
        & S-wave velocity  & $c_s$ & $\SI{3.464}{km/s}$ \\
        & Density  & $\rho$ & $\SI{2670}{kg/m^3}$ \\
        \hline
        Frictional Properties &  &  & \\
               
        & Steady  state  friction  coefficient  & $f_0$ & $\SI{0.6}{}$ \\

        & Reference  slip  velocity  & $V_0$ & $\SI{1}{\micro m/s}$ \\
        & Direct  effect  parameter  & $a$ & $\SI{0.01}{}$ \\
        & State  evolution effect parameter  & $b$ & $\SI{0.014}{}$ \\
        & Fully weakened friction coefficient & $f_w$ & $\SI{0.25}{}$ \\
        & Weakening slip velocity & $V_w$ & $\SI{0.15}{m/s}$ \\
        & State evolution distance  & $L$ & $\SI{0.4}{m}$ \\

        \hline
        TP Properties &  &  & \\
        & Thermal diffusivity  & $\alpha_\mathrm{th}$ & $\SI{1e-6}{m^2/s}$ \\
        & Volumetric heat capacity  & $\rho c$ & $\SI{2.7}{\mega J/m^3K}$ \\
        & Pore pressure increase per unit temperature  & $\Lambda$ & $\SI{0.1}{\mega Pa/K}$ \\
        & Hydraulic diffusivity  & $\alpha_\mathrm{hy}$ & ($15, 18, 21, 24, 27,$ \\ 
        & & & $  30, 33$) $\times 10^{-5} \, \SI{}{m^2/s}$ \\
        & Shear-zone half-width  & $w$ & $15, 18, 21, 24, 27,$ \\
        & & & $  30, 33$ \SI{}{mm} \\
       
        \\
        \hline
        \end{tabular}
        
    }
    
    \label{tab:MP-FP-TP-IC}
\end{table}

\newpage 
\section*{List of Figures} 


\newpage 


\newpage 


\newpage 


\newpage 


\newpage 


\newpage 


\newpage 


\newpage 


\newpage 


\newpage 


\newpage 
\title{Electronic Supplement \\~\\~\\ How does thermal pressurization of pore fluids affect earthquake dynamics and ground motions?}


\maketitle

\renewcommand\thefigure{S\arabic{figure}}    
\setcounter{figure}{0}    



\newpage
This electronic supplement contains temporal evolution of slip rate, total traction, temperature and pressure at different on-fault locations. It also describes effects of hydraulic diffusivity and shear-zone half-width variations on moment magnitude. The details of initial shear stress and rupture nucleation method are also provided here. 


\newpage 
\begin{figure}[H]
    \centering
    \includegraphics[width=1.0\textwidth]{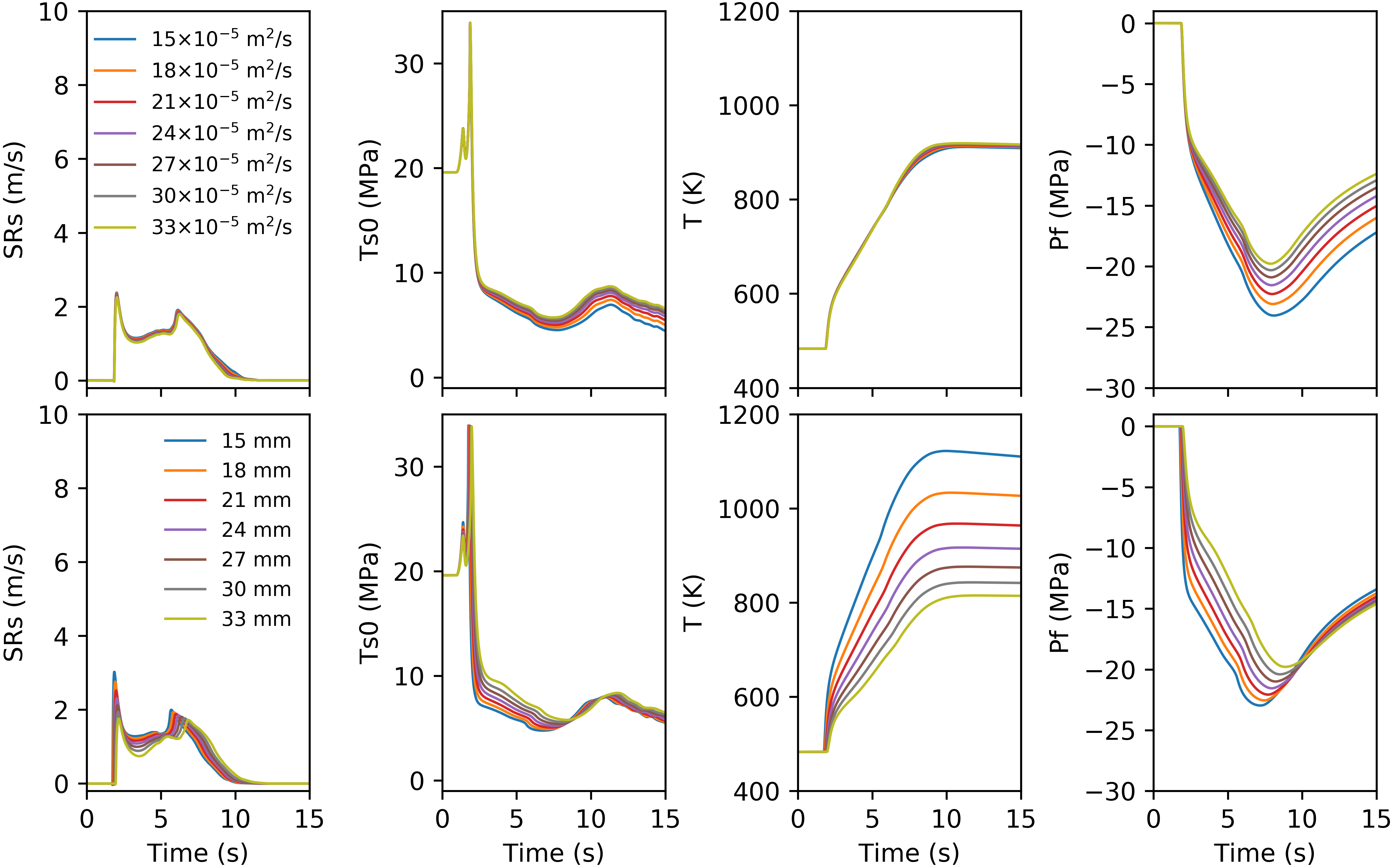}
    \caption{Temporal evolution of along strike slip rate (SRs), shear traction along strike (Ts0), temperature (T) and fluid pressure (Pf) at receiver s1 (Figure \ref{fig:rec}) for varying $\alpha_\mathrm{hy}$ and $w$ (color-coded)}
    \label{fig:s1-SRs}
\end{figure}


\newpage 
\begin{figure}[H]
    \centering
    \includegraphics[width=1.0\textwidth]{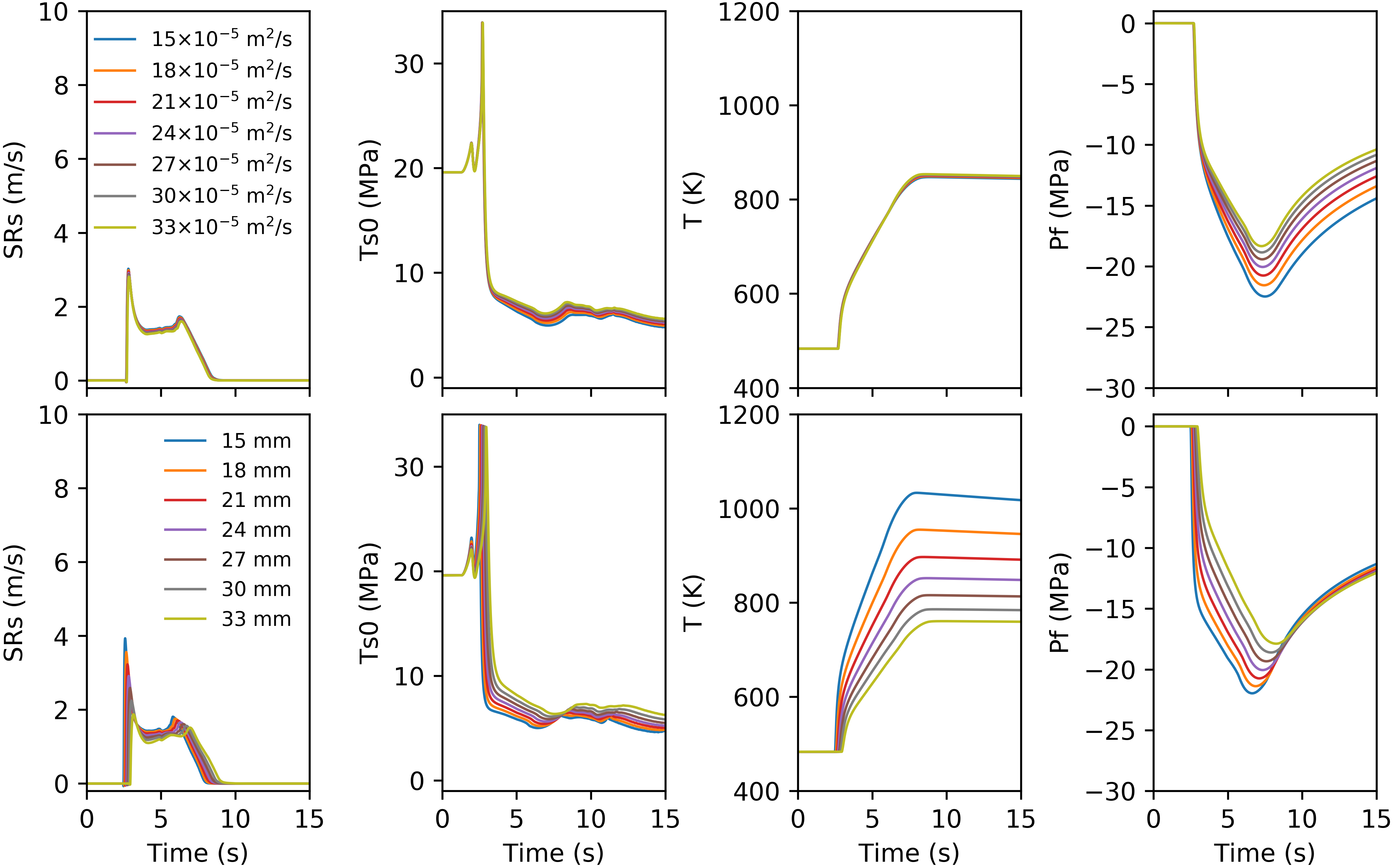}
    \caption{Temporal evolution of along strike slip rate (SRs), shear traction along strike (Ts0), temperature (T) and fluid pressure (Pf) at receiver s3 (Figure \ref{fig:rec}) for varying $\alpha_\mathrm{hy}$ and $w$ (color-coded).}
    \label{fig:s3-SRs}
\end{figure}


\newpage 
\begin{figure}[H]
    \centering
    \includegraphics[width=1.0\textwidth]{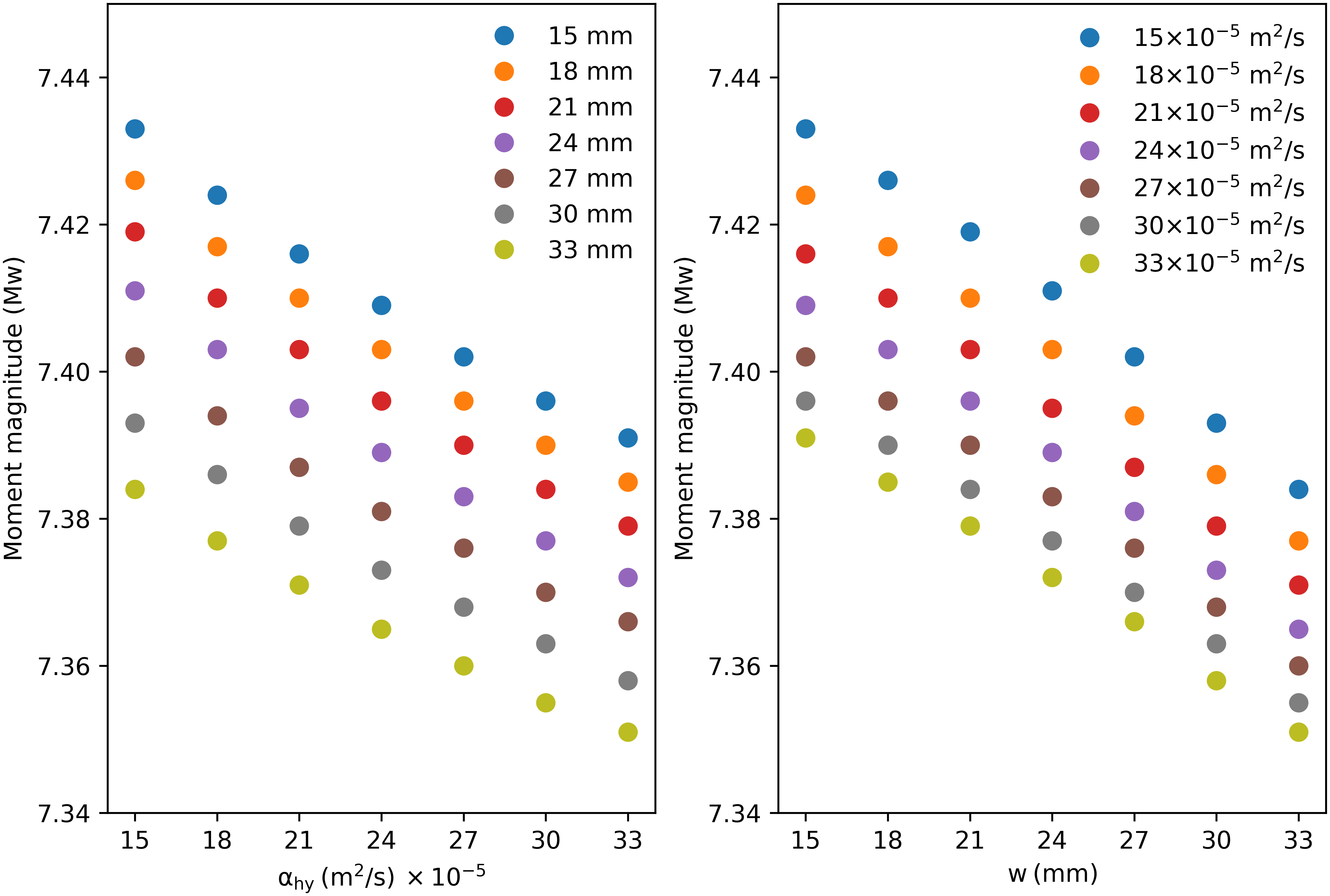}
    \caption{Variations of modeled moment magnitude $M_w$ with $\alpha_\mathrm{hy}$ and $w$ (color-coded).}
    \label{fig:Mw-var}
\end{figure}


\appendix
\renewcommand{\theequation}{\thesection.\arabic{equation}}
\newpage 
\section{Initial Conditions} \label{appendix:IniCond}
\setcounter{equation}{0}

\paragraph{}
The initial shear stress on the fault is $\tau_{\mathrm{ini}}(x,z)$, the effective normal  stress  is $\overline \sigma_{\mathrm{ini}}(x,z)$, the  initial  value  of  the  state  variable  is $\Psi_{\mathrm{ini}}(x,z)$ and initial fault sliding velocity is $V=V_{\mathrm{ini}}$ (= \SI{1e-16}{\m\per\s}). 

The initial normal stress and shear stress (in Pa) are given by:

\begin{equation}
    \overline \sigma_{\mathrm{ini}}(x,z) = \min((\rho-\rho_w)gz,\,  40 \times 10^6)
\end{equation}
\begin{equation}
    \tau_{\mathrm{ini}}(x,z) = 0.49 \times \overline\sigma_{\mathrm{ini}}(x,z)
\end{equation}

with $\rho_w$ = 1000 kg/$m^3$ the water density and $g$ = 9.8 $m/s^2$ the gravitational acceleration.

The initial value of the state variable $\Psi$  is given by:
\begin{equation}
    \Psi_{\mathrm{ini}}(x,z) = a\ln{\left[\frac{2V_0}{V_{\mathrm{ini}}}\sinh {\left(\frac{\tau_{\mathrm{ini}}}{a\overline \sigma_{\mathrm{ini}}}\right)}\right]} 
\end{equation}
with $V_0 = \SI{1}{\micro m/s}$ the reference  slip  velocity and $a = 0.01$ the direct effect parameter.

\section{Nucleation Method} \label{appendix:NucProc}
\setcounter{equation}{0}

\paragraph{}
The rupture is nucleated  by  imposing  a  horizontal shear  traction  perturbation that  depends  on  both  space  and time. The  perturbation  smoothly  grows  from  zero  to  its maximum  amplitude $\Delta \tau_0$ over  a  finite  time  interval $T_\textsc{nuc}$ and is confined to a finite region of the fault of radius R. 
Specifically, the perturbation is
\begin{equation}
    \Delta\tau(x,z,t) = \Delta \tau_0 F \left(\sqrt{(x-x_0)^2+(z-z_0)^2} \right) G(t),
\end{equation}
in which
\begin{equation}
    F(r) =
    \begin{cases}
        \exp \left(\frac{r^2}{r^2-R^2} \right), \,\, r<R
        \\
        0, \,\, r \ge R
    \end{cases}
\end{equation}
and
\begin{equation}
    G(t) =
    \begin{cases}
        \exp \left(\frac{(t-T_\textsc{nuc})^2}{t(t-2T_\textsc{nuc})} \right), \,\, 0<t<T
        \\
        1, \,\, t \ge T
    \end{cases}
\end{equation}
The perturbation is radially symmetric, with the radial distance away from the hypocenter along  the  fault  given  by
$r = \sqrt{(x-x_0)^2+(z-z_0)^2}$.
The  nucleation  parameters  are  given in the table below.
\begin{table}[h]
    \centering
    \begin{tabular}{|c|c|c|c|}
    \hline
    $\Delta \tau_0$ & $R$ & 
    $T_\textsc{nuc}$ & $(x_0,z_0)$ 
    \\
    \hline
    \SI{45}{\mega\Pa} & \SI{1.5}{\kilo\m} & \SI{1}{\s} & (\SI{-4}{\kilo\m}, \SI{-7.5}{\kilo\m})
    \\
    \hline
    \end{tabular}
\end{table}

\end{document}